\documentclass[12pt,a4paper,english,superscriptaddress,aps,nofootinbib]{revtex4}
\usepackage[utf8]{inputenc}
\usepackage[T1]{fontenc}
\usepackage{amsmath,amssymb,graphicx}
\makeatletter
\usepackage{babel}
\usepackage[active]{srcltx}
\usepackage{graphicx,color}
\usepackage{subfigure}
\usepackage{changebar}
\usepackage{upgreek}
\usepackage{hyperref}
\hypersetup{colorlinks=true,urlcolor= blue,citecolor=blue,linkcolor= blue}
\usepackage{braket}
\usepackage{dsfont}
\usepackage{mathtools}
\usepackage{slashed}
\usepackage{empheq}
\usepackage{tikz}
\usepackage{multirow}
\usetikzlibrary{decorations.pathmorphing}

\usepackage{graphicx}
\usepackage{amsfonts}
\usepackage{amsmath}
\newcommand{\be}{\begin{equation}}
	\newcommand{\ee}{\end{equation}}
\newcommand{\bea}{\begin{eqnarray}}
	\newcommand{\eea}{\end{eqnarray}}

\begin{document}

\title{Geodesics and Thermodynamics of a Schwarzschild Black Hole with Hernquist Dark Matter}

\author{A. R. P. Moreira}
\email[]{E-mail: allan.moreira@fisica.ufc.br}
\affiliation{Secretaria da Educa\c{c}\~ao do Cear\'a (SEDUC), Coordenadoria Regional de Desenvolvimento da Educa\c{c}\~ao (CREDE 9), Horizonte, Cear\'a, Brazil.}

\author{A. Bouzenada}
\email[]{E-mail: abdelmalekbouzenada@gmail.com}
\affiliation{Laboratory of Theoretical and Applied Physics, Echahid Cheikh Larbi Tebessi University, Tebessa 12001, Algeria}
\affiliation{Research Center of Astrophysics and Cosmology, Khazar University, Baku AZ1096, 41 Mehseti Street, Azerbaijan}

\author{F. M. Belchior}
\email[]{E-mail: belchior@fisica.ufc.br}
\affiliation{Departamento de Física, Universidade Federal da Paraíba, Centro de Ciências Exatas e da Natureza, João Pessoa, Paraíba, Brazil}

\author{F. C. E. Lima}
\email[]{E-mail: cleiton.estevao@ufabc.edu.br (corresponding author)}
\affiliation{Centro de Matématica, Computação e Cognição (CMCC), Universidade Federal do ABC (UFABC), Av. dos Estados 5001, Santo Andr\'{e}, S\~{a}o Paulo, Brazil.}

\begin{abstract}
In this work, we investigate the physical and geometrical properties of a Schwarzschild black hole (BH) immersed in a Hernquist dark matter halo. To accomplish our purpose, one builds the BH metric by incorporating the Hernquist dark matter profile into the Schwarzschild geometry. In addition, we verify the null geodesic solutions and the Halo effect on photon dynamics. Within this framework, one examines the corresponding light trajectories to determine the deformation of photon paths generated by the dark matter distribution. Furthermore, the thermodynamic properties of the system are studied by deriving expressions for the black hole mass, the horizon condition, the Hawking temperature, the entropy, the Gibbs free energy, and the heat capacity. Our results show that the dark matter halo modifies the thermal structure and stability conditions of the black hole configuration. Finally, we investigate the scalar perturbations to examine the influence of the Hernquist halo on the dynamical propagation of scalar fields in the BH background. In this framework, the results obtained demonstrate that the dark matter parameters yield nontrivial corrections to the optical, thermodynamic, and perturbative properties of the Schwarzschild black hole, producing deviations from the standard vacuum solution.
\end{abstract}

\maketitle

\section{Introduction}\label{sec:1}

Black holes (BHs) are among the most remarkable predictions of Einstein's gravitational field equations, formulated in 1916~\cite{GR1} within the framework of General Relativity (GR)~\cite{GR2}. In essence, they are regions of spacetime where gravity becomes so intense that no physical signal that not even light can escape. This defining feature, the event horizon (EH), establishes a sharp causal boundary between the BH interior and the surrounding universe. Because of the extreme spacetime curvature involved, BHs offer a natural laboratory for probing relativistic gravitation, high-energy astrophysical processes, and quantum field theory in curved backgrounds.
 
For much of the twentieth century, BHs remained largely theoretical constructs. That changed dramatically in 2015, when the Laser Interferometer Gravitational-Wave Observatory (LIGO) detected gravitational waves (GWs) generated by the merger of two stellar-mass BHs~\cite{BH1}, which was the first direct observational confirmation of GR operating in the strong-field regime, and the opening of an entirely new branch of observational astronomy. A few years later, in 2019, the Event Horizon Telescope (EHT) collaboration released the first horizon-scale image of the supermassive BH at the center of the galaxy $M87^{*}$, a bright and asymmetric emission ring encircling a dark central shadow, in close agreement with GR predictions~\cite{BH2,BH3,BH4,BH5,BH6,BH7}. Subsequent polarimetric observations of the same source provided quantitative information on the magnetic field structure near the horizon, lending support to magnetically arrested accretion models and shedding light on the mechanisms driving relativistic jet formation~\cite{BH8,BH9}. More recently, the EHT turned its focus to Sagittarius~$A^{*}$ (Sgr~$A^{*}$), the supermassive BH at the Galactic center, finding that the observed shadow radius and angular profile are again consistent with GR expectations, despite the considerably more variable accretion environment~\cite{BH10,BH11}. In this sense, these results involving spanning gravitational-wave measurements and direct imaging have established BHs as astrophysical objects that can be studied observationally, opening the door to tests of strong-field gravity, relativistic plasma dynamics, and the connection between spacetime curvature and energetic processes~\cite{BH12,BH13,BH14,BH15,BZ1,BZ2,BZ3,BZ4,BZ5,BZ6,BZ7,BZ8,BZ9,BZ10,BZ11}.
 
The case for dark matter (DM) rests on a broad and mutually independent body of observational evidence: the anomalously flat rotation curves of spiral galaxies~\cite{DH1,DH2,DH3}, the gravitational lensing signature of colliding clusters such as the Bullet Cluster~\cite{DH4}, and baryon acoustic oscillations extracted from large-scale structure surveys~\cite{DH5}. None of these datasets can be reconciled within standard gravitational theory without invoking a dominant non-baryonic matter component that contributes substantially to the total galactic and cosmological mass budget. In this picture, the supermassive BHs found at galactic centers reside inside extended DM halos governed by cold dark matter distributions~\cite{DH6,DH7}, so that the ambient DM medium naturally modifies the local spacetime geometry and, in turn, several physical observables in the strong-gravity regime.
 
To make this problem tractable, a number of analytical halo models have been proposed in the literature~\cite{DH8,DH9,DH10,DH11}. Among them, the Dehnen profile~\cite{DH9} stands out for its versatility: it is a spherically symmetric density distribution characterized by three parameters $(\alpha,\beta,\gamma)$ that independently control the inner and outer logarithmic slopes of the halo. Depending on the choice of these parameters, the model can describe both cored and cuspy density profiles, making it a flexible template for studying generic DM contributions to compact gravitational systems. A physically important special case is the Hernquist profile~\cite{DH12}, recovered for $(\alpha,\beta,\gamma)=(1,4,1)$, which reproduces the observed mass distribution of elliptical galaxies and admits a fully analytical gravitational potential. The Hernquist profile has recently been used to construct a Schwarzschild BH embedded in a DM halo, and the resulting geometry was analyzed in terms of its thermodynamic properties, weak gravitational lensing, and observational constraints~\cite{DH13}. In parallel, Feng and Zhang~\cite{DH14} studied the shadow and quasi-normal mode (QNM) spectrum of this Schwarzschild–Hernquist spacetime, showing that the halo compactness shifts the shadow radius and introduces a redshift in the QNM frequencies. Despite these advances, several strong-field gravitational effects of this geometry remain unexplored, particularly the structure of the lensing ring and photon ring in multi-image configurations, as well as independent determinations of the QNM spectrum using complementary numerical and analytical approaches.
 
The nature of DM is one of the central open questions in theoretical physics~\cite{b1}. Multiple observational probes — galaxy rotation dynamics, large-scale structure formation, gravitational lensing in clusters, and precision measurements of the cosmic microwave background (CMB) — all point to the same conclusion: a substantial fraction of the Universe's energy content is invisible to electromagnetic observations~\cite{b2,b3,b4,b5,b6,b7}. Primordial nucleosynthesis combined with CMB data constrains the DM energy fraction to be roughly one quarter of the total, and this non-baryonic component is essential for seeding the density perturbations that eventually collapse into the galaxies and galaxy clusters we observe today. While certain phenomena can be accommodated within modified gravity frameworks, several key observational signatures remain incompatible with these alternatives, motivating instead a particle physics interpretation in which DM consists of one or more species that are weakly or non-interacting and lie beyond the Standard Model~\cite{b8,b9,b10,b11,b12,b13,b14,b15}. Viable candidates are generally expected to be electrically neutral and colorless, consistent with the absence of electromagnetic emission or absorption that defines their dark character. The persistent null results from direct detection experiments further restrict possible couplings to Standard Model gauge sectors and reinforce the picture of a particle that is stable over cosmological timescales. Models with significant DM self-interactions are also discussed in the literature, though they are subject to stringent bounds from halo morphology and galaxy cluster dynamics, which impose upper limits on the elastic scattering cross section~\cite{b16,b17,b18}.
 
On galactic scales, once a DM halo has grown sufficiently massive, its structure is governed primarily by the gravitational potential of the host system. The nonlinear evolution and complex merger history of halos make an exact determination of their density profile generally nontrivial; in practice, the profile shape is treated as an effective quantity constrained by observational data~\cite{BZ12,BZ13,BZ14,BZ15,BZ16,BZ17,BZ18,BZ19,BZ20,BZ21}. The spatial extent of the halo depends jointly on the DM particle mass $m$ and the gravitational environment of the host object. For planetary systems like Earth, a DM particle mass below roughly $10^{-9}$~eV yields a halo that extends beyond the planetary radius and is therefore potentially accessible to surface or near-orbital detectors~\cite{b19}. In the opposite limit, $m \gg 10^{-9}$~eV, the halo contracts to scales well below the Earth's radius, greatly reducing its observational accessibility.
 
The thermodynamic interpretation of BHs traces back to the identification of horizon area with entropy and to the prediction of thermal radiation emitted by event horizons, established in the seminal works of Bekenstein and Hawking~\cite{BT1,BT2}. This correspondence provides a precise mapping between gravitational configurations and the laws of standard thermodynamics, and it serves as a crucial testing ground for quantum gravity programs in which horizon dynamics are attributed a statistical-mechanical origin. Quantum field theory in curved spacetime~\cite{BT3} and semiclassical path-integral methods~\cite{BT4,BT5} both confirm that BHs satisfy formal analogues of the four laws of thermodynamics, with well-defined temperature, entropy, and equilibrium states, while Hawking radiation arises as a consequence of vacuum fluctuations near the horizon.
 
Beyond this formal structure, BH systems exhibit a rich thermodynamic phase structure whose form depends on boundary conditions and conserved charges. For the asymptotically anti-de Sitter (AdS) Schwarzschild BH, small and large BH branches emerge, and the transition between thermal radiation and large BHs is governed by the Hawking–Page transition~\cite{BT6}, which can be interpreted as a change in the dominant saddle point of the gravitational partition function. For charged solutions, such as Reissner–Nordström–AdS BHs, the phase diagram becomes considerably richer, displaying small, intermediate, and large BH phases with critical points and coexistence curves that closely parallel those of van der Waals fluids~\cite{BT7,BT8,BT9}. The area–entropy relation itself points toward an underlying microstate structure associated with spacetime geometry, imposing non-trivial consistency conditions on any candidate theory of quantum gravity~\cite{BT10,BT11}. It has also played a central role in the formulation of holographic principles and gauge/gravity duality, where higher-dimensional gravitational theories are mapped to lower-dimensional quantum field theories~\cite{BT12,BT13,BT14,BT15,BT16}. Within this dual description, the Hawking–Page transition acquires an interpretation as a confinement–deconfinement transition in the boundary gauge theory~\cite{BT17}. More recently, motivated by holographic ideas, BH thermodynamics has been extended to an enlarged phase space in which the cosmological constant plays the role of thermodynamic pressure, with a conjugate volume, and the first law is enriched by additional thermodynamic variables and chemical potentials. This analysis is a framework that has led to a refined understanding of critical phenomena and phase transitions in gravitational systems~\cite{BT18,BT19,BT20,BT21,BT22,BT23}.
 
The rest of this paper is organized as follows. In Sec.~(\ref{S2}), we introduce the Schwarzschild BH surrounded by a Hernquist DM halo, presenting the spacetime geometry, metric function, and physical parameters that define the gravitational background. Section~(\ref{S3}) examines the geometric and physical properties of the model: the effective potential for null geodesics and its dependence on the BH and DM parameters are analyzed in Sec.~(\ref{S3-1}), with particular limiting cases discussed in Sec.~(\ref{S3-1-1}), while photon trajectories and gravitational lensing properties are studied in Sec.~(\ref{S3-2}), with additional limiting configurations treated in Sec.~(\ref{S3-2-1}). The thermodynamic behavior of the system is the subject of Sec.~(\ref{S4}): the horizon structure and mass relation are derived in Sec.~(\ref{S4-1}), the Hawking temperature is computed in Sec.~(\ref{S4-2}), the entropy is discussed in Sec.~(\ref{S4-3}), the Gibbs free energy and phase structure are analyzed in Sec.~(\ref{S4-4}), and the heat capacity together with thermal stability criteria are investigated in Sec.~(\ref{S4-5}). Section~(\ref{S5}) is devoted to scalar perturbations of the spacetime, with several particular configurations examined in Sec.~(\ref{S5-1}). Finally, our main results and their physical implications are summarized and discussed in Sec.~(\ref{S6}).

\section{BH Model Metric}\label{S2}

Let us start our study by adopting a static and spherically symmetric BH spacetime embedded in a Hernquist dark matter halo distribution, whose spacetime line element is
\begin{align}
    ds^{2}=-f(r)\,dt^{2}+\frac{dr^{2}}{f(r)}+r^{2}\,d\Omega^{2},
    \label{aa1}
\end{align}
with
\begin{align}
    f(r)=\exp\left(-\frac{4\pi r_{s}^{3}\rho_{s}}{r+r_{s}}\right)-\frac{2M}{r}-\alpha \qquad \mathrm{and} \qquad d\Omega^{2}=d\theta^{2}+\sin^{2}\theta\, d\phi^{2}.
\label{aa2}
\end{align}
Here, the parameter $M$ denotes the gravitational mass of the BH, $\rho_{s}$ and $r_{s}$ are, respectively, the characteristic density and scale radius of the Hernquist dark matter profile \cite{DH12,DH6,DH7,DH9}. Meanwhile, the constant parameter $\alpha$ describes an additional geometric contribution modifying the background spacetime configuration \cite{BZ1,BZ2,BZ3,BZ12,BZ13}.

The metric function in Eq. \eqref{aa2} couples the Schwarzschild term to the correction generated by the surrounding dark matter distribution \cite{DH1,DH2,DH4,b1,b13}. The exponential factor modifies the vacuum spacetime geometry by determining the gravitational contribution of the Hernquist halo to the BH background \cite{DH10,DH11,DH12}. Thus, one notes that the spacetime structure depends explicitly on the parameters $(\rho_{s},r_{s},\alpha)$, which govern the dark matter density distribution, halo scale length, and geometric deformation \cite{BZ4,BZ5,BZ6,BZ14,BZ15}.

The condition $f(r_h)=0$ defines the event horizon radius, which brings us to
\begin{align}
    \exp\left(-\frac{4\pi r_{s}^{3}\rho_{s}}{r_{h}+r_{s}}\right)-\frac{2M}{r_{h}}-\alpha=0.
\label{aa3}
\end{align}
This relation shows that the event horizon receives direct corrections from the dark matter halo and constant ($\alpha$) parameters \cite{BZ7,BZ8,BZ9}. Therefore, unlike the Schwarzschild spacetime, the horizon radius is not described solely by the BH mass parameter \cite{GR2,BH1,BH2}.

Several limiting configurations follow from the metric function (\ref{aa2}). For instance, in the absence of dark matter density, one has $\rho_s\to 0$ and
\begin{align}
\exp\left(-\frac{4\pi r_{s}^{3}\rho_{s}}{r+r_{s}}\right)\to 1,
\end{align}
which results in
\begin{align}
    f(r)=1-\frac{2M}{r}-\alpha.
    \label{aa4}
\end{align}
Therefore, for the additional limit $\alpha\to 0$, the standard Schwarzschild BH geometry is recovered \cite{GR1,GR2}, i.e.,
\begin{align}
    f(r)=1-\frac{2M}{r}.
    \label{aa5}
\end{align}
Accordingly, our model is a Schwarzschild-like spacetime extension coupled to the Hernquist dark matter halo and parameter $\alpha$ contribution \cite{DH12,BZ10,BZ11}.

Meanwhile, in the large-distance regime $(r\gg r_{s})$, the exponential factor boils down to
\begin{align}
    \exp\left(-\frac{4\pi r_{s}^{3}\rho_{s}}{r+r_{s}}\right)\simeq 1-\frac{4\pi r_{s}^{3}\rho_{s}}{r+r_{s}}+\mathcal{O}\left(\rho_{s}^{2}\right),
\end{align}
which lead us to
\begin{align}
    f(r)\simeq 1-\alpha-\frac{2M}{r}-\frac{4\pi r_{s}^{3}\rho_{s}}{r+r_{s}}.
    \label{aa6}
\end{align}
This expression shows that the dark matter halo contributes an additional attractive term to the gravitational field in the asymptotic region. Consequently, the spacetime geometry departs from the Schwarzschild vacuum solution due to the combined contributions of the halo density and the scale parameter $r_{s}$.

In the near-center region $(r\ll r_{s})$, the metric function takes the form
\begin{align}
f(r)\simeq \exp\left(-4\pi r_{s}^{2}\rho_{s}\right)-\frac{2M}{r}-\alpha,
\label{aa7}
\end{align}
which shows that the dark matter distribution modifies the local gravitational structure through a constant exponential correction factor. Furthermore, one notes that the modification becomes stronger when the values of $\rho_{s}$ and $r_{s}$ increase.

Relative to the Schwarzschild BH immersed in a Hernquist dark matter halo, our model has an additional parameter $\alpha$, which introduces extra corrections to the spacetime geometry. Also, it is essential to highlight that the Schwarzschild-Hernquist configuration is recovered from Eq. (\ref{aa2}) in the limit $\alpha=0$.

\begin{figure}[ht!]
\begin{center}
\begin{tabular}{ccc}
\includegraphics[height=4cm,width=5cm]{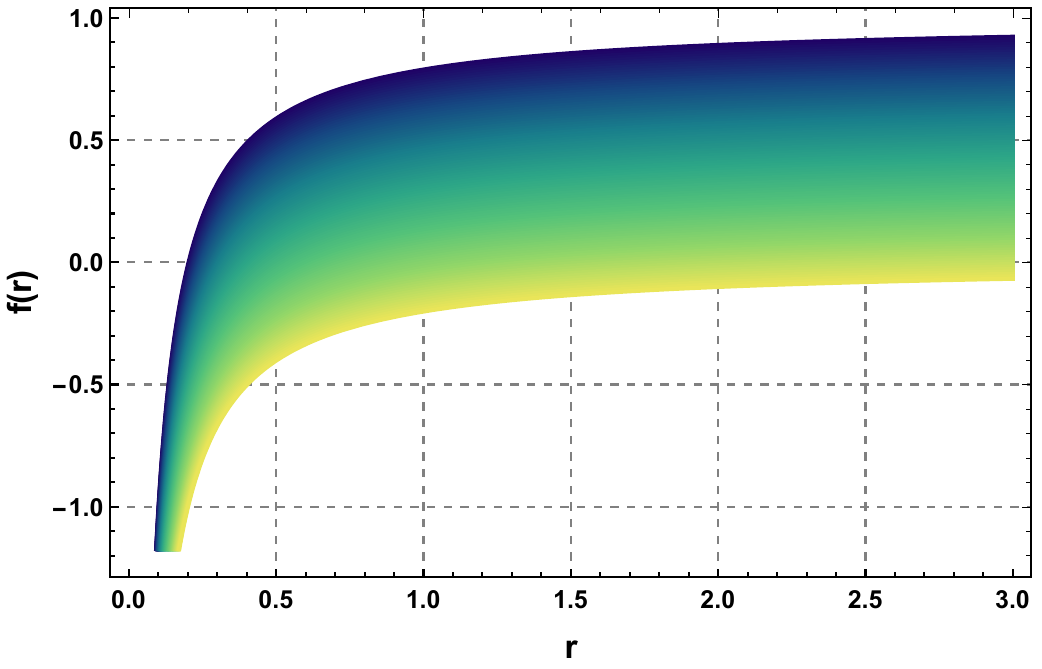} 
\includegraphics[height=4cm]{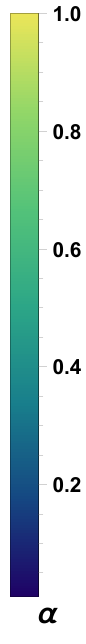}
\includegraphics[height=4cm,width=5cm]{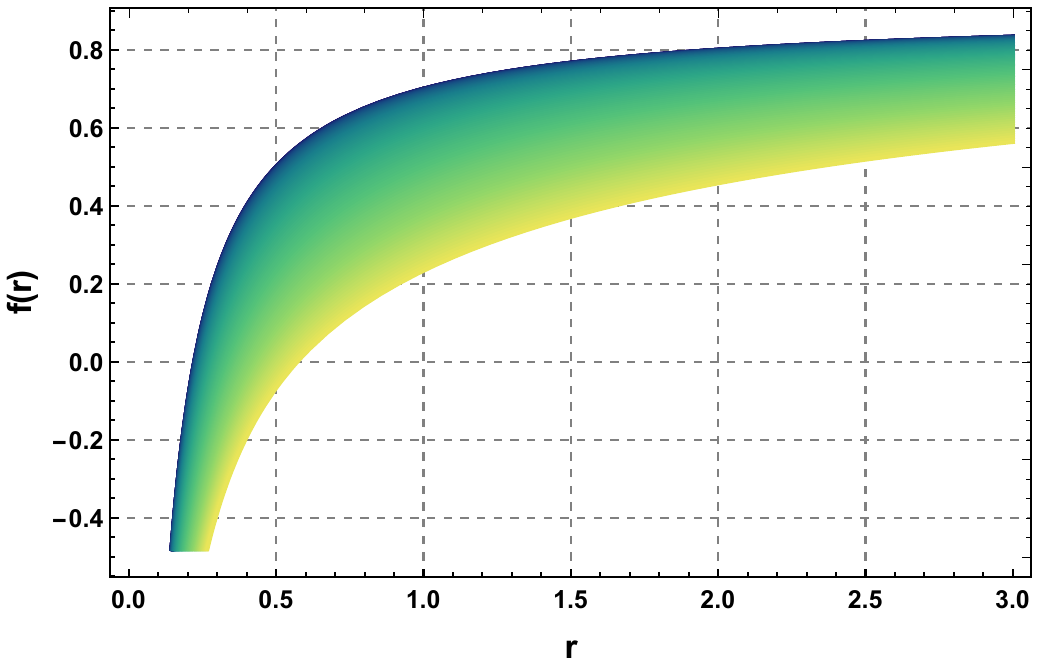} 
\includegraphics[height=4cm]{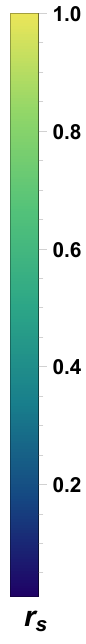}
\includegraphics[height=4cm,width=5cm]{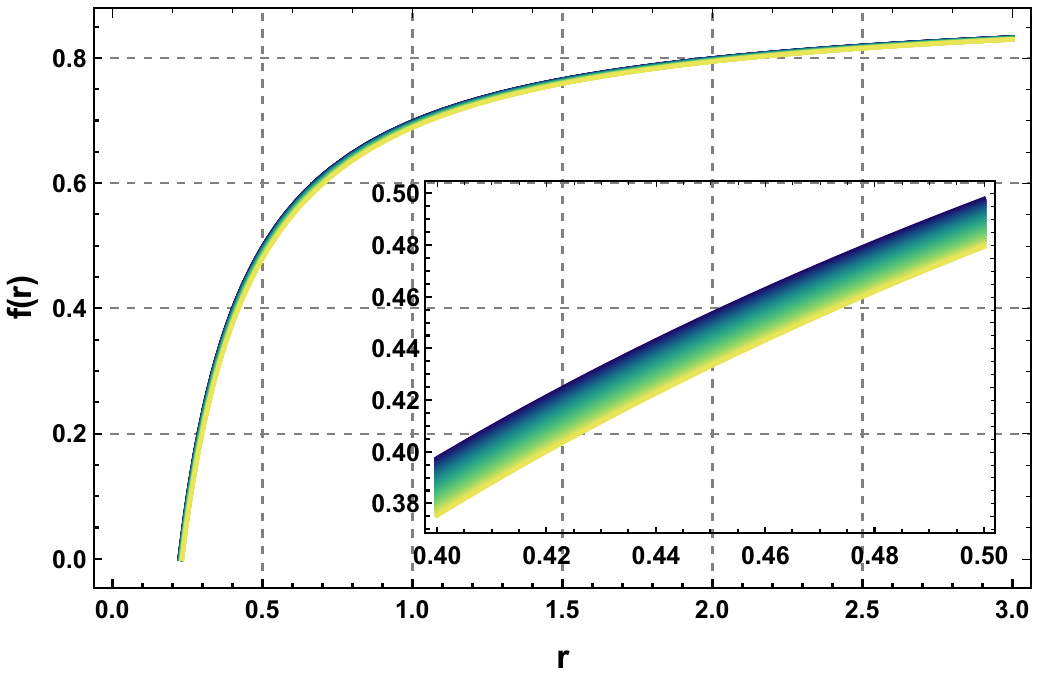} 
\includegraphics[height=4cm]{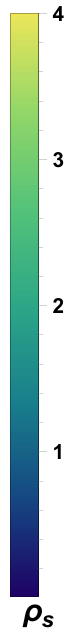}\\
\hspace{0.5cm} (a) $r_s=\rho_s=0.5$ \hspace{2.5cm} (b) $\alpha=\rho_s=0.5$ \hspace{3cm} (c) $r_s=\alpha=0.5$\\
\end{tabular}
\end{center}
\caption{Profile of the metric function $f(r)$ vs. $r$ with $M=0.1$.\label{fig01}}
\end{figure}

In this framework, the parameter $\alpha$ determines deviations from the conventional Schwarzschild-Hernquist BH, modifying the horizon configuration, effective potential, and gravitational characteristics of the spacetime. Besides, we highlighted that the combined contributions of the parameters $\rho_{s}$, $r_{s}$, and $\alpha$ produce nontrivial corrections to the Schwarzschild BH geometry.
We exposed in Figure~\ref{fig01} the behavior of the metric function $f(r)$ for different values of the parameters $(\alpha,r_s,\rho_s)$ associated with the Hernquist dark matter distribution and the geometric deformation of the spacetime. In Fig. \ref{fig01}(a), for fixed $r_s=\rho_s=0.5$, increasing values of the parameter $\alpha$ produce a global downward shift of the metric function, reducing the asymptotic value of $f(r)$ and modifying the location of the event horizon. This behavior indicates that $\alpha$ strengthens the effective gravitational attraction, expanding the region where strong-field effects become dominant. In Fig. \ref{fig01}(b), for fixed $\alpha=\rho_s=0.5$, variations of the scale radius $r_s$ significantly alter the transition of the metric function between the near-horizon and asymptotic regions. Larger values of $r_s$ smooth the radial evolution of the geometry, reflecting the extended influence of the Hernquist halo over the spacetime structure. Finally, we show in Fig. \ref{fig01}(c) the effect of the density parameter $\rho_s$ for fixed $r_s=\alpha=0.5$. 

The results show us that the increase in $\rho_s$ modifies the exponential contribution associated with the dark matter distribution, providing visible corrections in the intermediate and asymptotic regimes. Physically, these results demonstrate that the combined action of the parameters $(\rho_s,r_s,\alpha)$ generates nontrivial deformations of the Schwarzschild geometry, directly affecting the horizon structure, the effective gravitational potential, and consequently the optical, dynamical, and thermodynamic properties of the black hole spacetime.

\section{Geometric and Physical Properties}\label{S3}

Null geodesics describe the propagation of massless particles such as photons in curved spacetime \cite{GR1,GR2}. In the vicinity of a BH, these trajectories determine the optical response from the geometry, including gravitational lensing, the existence of photon spheres, and the shadow structure. Mathematically, null geodesics satisfy the condition
\begin{align}
    g_{\mu\nu}\,\frac{dx^{\mu}}{d\lambda}\,\frac{dx^{\nu}}{d\lambda} = 0,
\end{align}
where $\lambda$ is the affine parameter along the trajectory \cite{GR2}. 

\subsection{Effective Potential for Null Geodesics}\label{S3-1}

Let us consider the motion of massless particles in the equatorial plane defined by $\theta=\pi/2$ and $\dot{\theta}=0$. Thus, the line element \eqref{aa1} for the metric function \eqref{aa2} takes the form
\begin{align}
    ds^2=-\left[\exp\left(-\frac{4\pi r_s^{3}\rho_s}{r+r_s}\right)-\frac{2M}{r}-\alpha\right]\,dt^2 + \frac{dr^2}{\left[\exp\left(-\frac{4\pi r_s^{3}\rho_s}{r+r_s}\right)-\frac{2M}{r}-\alpha\right]}+r^2\,d\phi^2.
\end{align}

Therefore, the corresponding Lagrangian density is
\begin{align}
    \mathcal{L}=\frac{1}{2}\,g_{\mu\nu}\,\frac{dx^{\mu}}{d\lambda}\,\frac{dx^{\nu}}{d\lambda}=\frac{1}{2}\left[-f(r)\left(\frac{dt}{d\lambda}\right)^2+\frac{1}{f(r)}\left(\frac{dr}{d\lambda}\right)^2+r^2\left(\frac{d\phi}{d\lambda}\right)^2\right].
\label{lagrangian_null}
\end{align}

Since $\mathcal{L}$ does not explicitly depend on $t$ and $\phi$, the energy $E$ and angular momentum $L$ are conserved. Within this framework, the energy and angular momentum are, respectively,
\begin{align}
    E=f(r)\,\frac{dt}{d\lambda} \qquad \mathrm{and} \qquad L=r^2\,\frac{d\phi}{d\lambda}.
    \label{constants_motion}
\end{align}

\begin{figure}[ht!]
\begin{center}
\begin{tabular}{ccc}
\includegraphics[height=4cm,width=5cm]{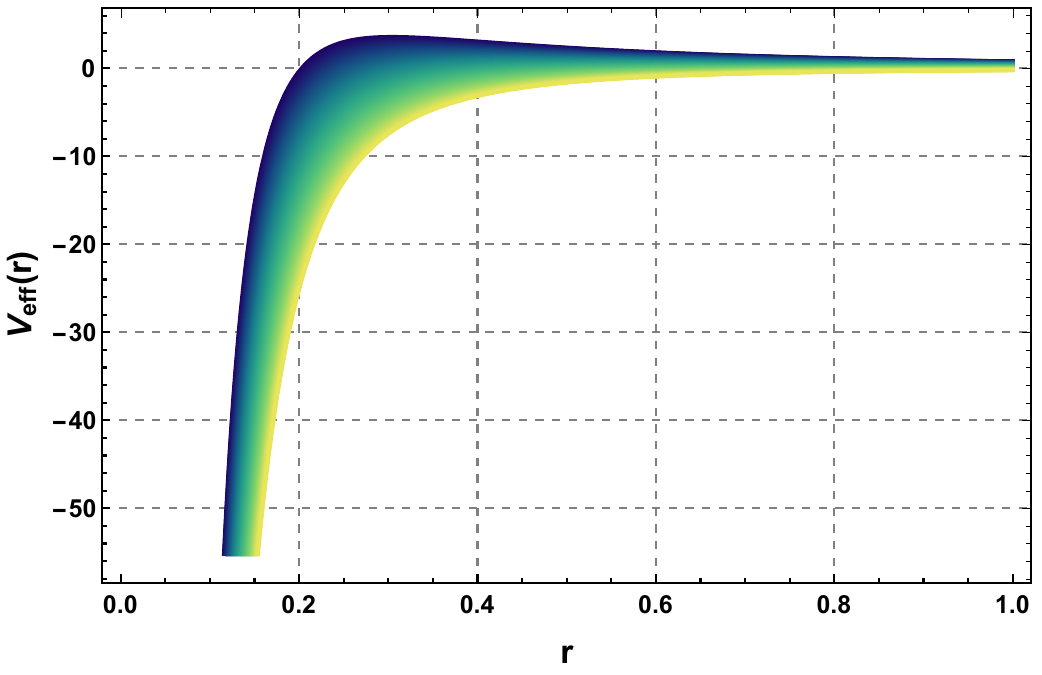} 
\includegraphics[height=4cm]{fig1aa.pdf}
\includegraphics[height=4cm,width=5cm]{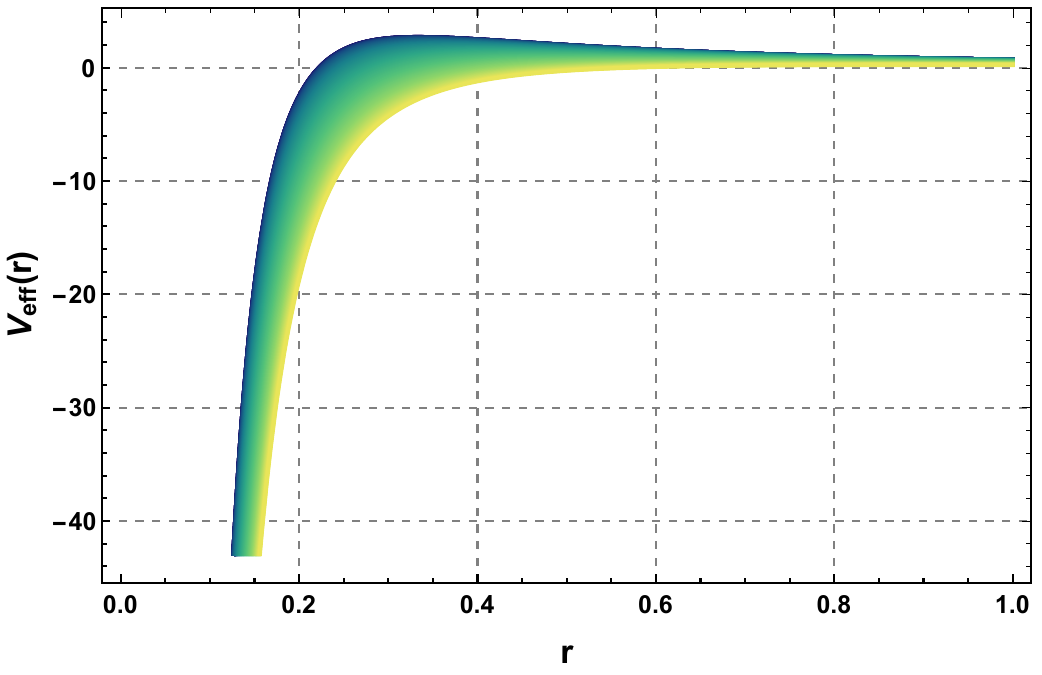} 
\includegraphics[height=4cm]{fig1bb.pdf}
\includegraphics[height=4cm,width=5cm]{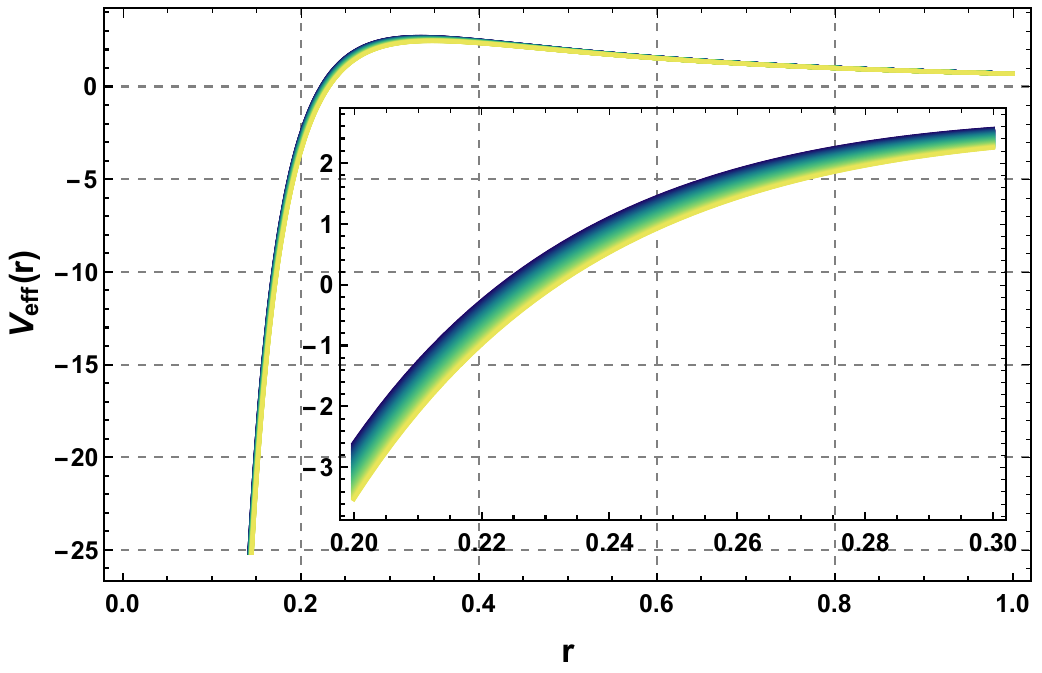} 
\includegraphics[height=4cm]{fig1cc.pdf}\\
\hspace{0.5cm} (a) $r_s=\rho_s=0.5$ \hspace{2.5cm} (b) $\alpha=\rho_s=0.5$ \hspace{3cm} (c) $r_s=\alpha=0.5$
\end{tabular}
\end{center}
\caption{The effective potential $V_{\mathrm{eff}}$ vs. $r$ with $M=0.1$.\label{fig02}}
\end{figure}

Substituting Eq.~\eqref{constants_motion} into Eq.~\eqref{lagrangian_null} and imposing the null constraint $\mathcal{L}=0$, one concludes that
\begin{align}
    \left(\frac{dr}{d\lambda}\right)^2 + V_{\text{eff}}(r) = E^{2},
\label{radial_null}
\end{align}
where the effective potential boils down to
\begin{align}
    V_{\text{eff}}(r)=\frac{L^{2}}{r^{2}}\left[\exp\!\left(-\frac{4\pi r_s^{3}\rho_s}{r+r_s}\right)-\frac{2M}{r}-\alpha\right].
    \label{veff_specific}
\end{align}
Naturally, this effective potential [see Figs. \ref{fig02}[(a)-(c)]encapsulates the influence of the exponential density profile characterized by $\rho_s$, $r_s$, and the constant deformation parameter $\alpha$. The structure of $V_{\text{eff}}(r)$ determines the existence and stability of circular photon orbits, which directly shape the optical appearance and BH shadow.

Figure \ref{fig02}[(a)-(c)] presents the behavior of the effective potential $V_{\text{eff}}(r)$ for null geodesics under different values of the parameters $(\alpha,r_s,\rho_s)$ that characterize the deformed BH spacetime surrounded by a Hernquist dark matter halo. In Fig. \ref{fig02}(a), with fixed values $r_s=\rho_s=0.5$, the increase of the deformation parameter $\alpha$ produces a lowering of the effective potential barrier, shifting its maximum toward smaller radial distances. Physically, this result indicates that the parameter $\alpha$ enhances the attractive force concerning the gravitational field, modifying the location and stability of unstable circular photon orbits. In Fig. \ref{fig02}(b), for fixed $\alpha=\rho_s=0.5$, variations of the scale parameter $r_s$ significantly affect both the width and the height of the potential barrier. Therefore, larger values of $r_s$ give rise to a broader potential structure, indicating that the spatial extension of the dark matter halo modifies photon propagation over a wider radial domain. Meanwhile, in Fig. \ref{fig02}(c), we expose the influence of the density parameter $\rho_s$ for fixed $r_s=\alpha=0.5$. An increase in $\rho_s$ enhances the height of the potential barrier, thereby indicating stronger gravitational confinement of null particles and a more pronounced photon-trapping mechanism.

\subsubsection{Particular Limits}\label{S3-1-1}

It is useful to analyze a few specific limiting cases of the metric function \eqref{aa2}, i.e., $\rho_s\to 0$, $\alpha\to 0$, and $r_s\to 0$. Let us start analyzing the condition $\rho_s\to 0$. Within this regime, the exponential term becomes unity and the metric function is
\begin{align}\label{fx}
    f(r)=1-\frac{2M}{r}-\alpha,
\end{align}
which leads us to the effective potential, 
\begin{align}
    V_{\text{eff}}(r)= \frac{L^{2}}{r^{2}}\left(1-\frac{2M}{r}-\alpha\right).
\end{align}
Note that, in the case $\rho_s = 0$, the exponential term becomes unity, reducing the metric function \eqref{fx} to a global monopole solution, see Ref. \cite{Barriola}. 

Meanwhile, when $\alpha=0$, the geometry is governed by the metric function, viz.,
\begin{align}
    f(r)=\exp\left(-\frac{4\pi r_s^{3}\rho_s}{r+r_s}\right)-\frac{2M}{r}.
\end{align}
Naturally, this case isolates the influence of the interior matter density profile on photon trajectories.

Finally, the case $r_s\to 0$, one notes that the exponential correction becomes trivial, i.e., 
\begin{align}
\exp\!\left(-\frac{4\pi r_s^{3}\rho_s}{r+r_s}\right)\longrightarrow 1,
\end{align}
and the metric reduces once again to the shifted Schwarzschild form.

These limits highlight how the exponential matter distribution and the constant parameter $\alpha$ alter the structure of null geodesics and, consequently, the shadow and lensing features of the BH spacetime.

\subsection{Light trajectories}\label{S3-2}

In this subsection, we investigate the propagation of photons in the spacetime determined by the metric function (\ref{aa2}). Thus, we analyze how its geometric and physical parameters affect the bending of light and the overall orbital structure of null geodesics.

Starting from the radial equation of motion for null geodesics, together with the effective potential definition (\ref{veff_specific}), the orbital equation for photons is
\begin{align}
\left(\frac{1}{r^{2}}\frac{dr}{d\phi}\right)^{2}
    =\frac{\mathrm{E}^{2}}{\mathrm{L}^{2}}-\frac{f(r)}{r^{2}}.
\label{orbit_general_new}
\end{align}
It is convenient to parametrize the energy-angular momentum ratio as $\frac{\mathrm{E}}{\mathrm{L}}=\frac{1}{b}$, where $b$ is the photon's impact parameter. Substituting the metric function~(\ref{aa2}) into~(\ref{orbit_general_new}), one obtains
\begin{align}
    \left(\frac{1}{r^{2}}\frac{dr}{d\phi}\right)^{2}=\frac{1}{b^{2}}-\frac{1}{r^{2}}\left[\exp\!\left(-\frac{4\pi r_s^{3}\rho_s}{r+r_s}\right)-\frac{2M}{r}-\alpha\right].
\label{orbit_r_new}
\end{align}

Introducing the inverse radial coordinate $u=1/r$, the expression (\ref{orbit_r_new}) transforms into
\begin{align}
\left(\frac{du}{d\phi}\right)^{2}
= \frac{1}{b^{2}}
- u^{2}\exp\!\left(-\frac{4\pi r_s^{3}\rho_s u}{1+r_s u}\right)
+ 2Mu^{3}
+ \alpha u^{2}.
\label{orbit_u_new}
\end{align}

By differentiating Eq. \eqref{orbit_u_new} with respect to $\phi$, one obtains the nonlinear differential equations that govern the photon trajectory
\begin{align}
    \frac{d^{2}u}{d\phi^{2}}+u=\frac{1}{2}\frac{d}{du}\left[u^{2}\exp\!\left(-\frac{4\pi r_s^{3}\rho_s u}{1+r_s u}\right)\right]-3 M u^{2}-\alpha u.
\label{orbit_diffeq_new}
\end{align}

Equation \eqref{orbit_diffeq_new} fully characterizes the orbital dynamics of light in this geometry. The exponential term, originating from the matter distribution encoded by the density parameter $\rho_s$ and scale $r_s$, introduces nontrivial modifications to the bending of light, particularly in the strong-field regime. The mass parameter $M$ governs the overall curvature, while the constant $\alpha$ acts as an additional deformation of the asymptotic structure of the spacetime. Together, these parameters determine the existence and stability of photon spheres, the deflection angle, and the qualitative features of the optical geometry around the compact object.
\begin{figure}[ht!]
\centering
\includegraphics[height=6cm,width=6cm]{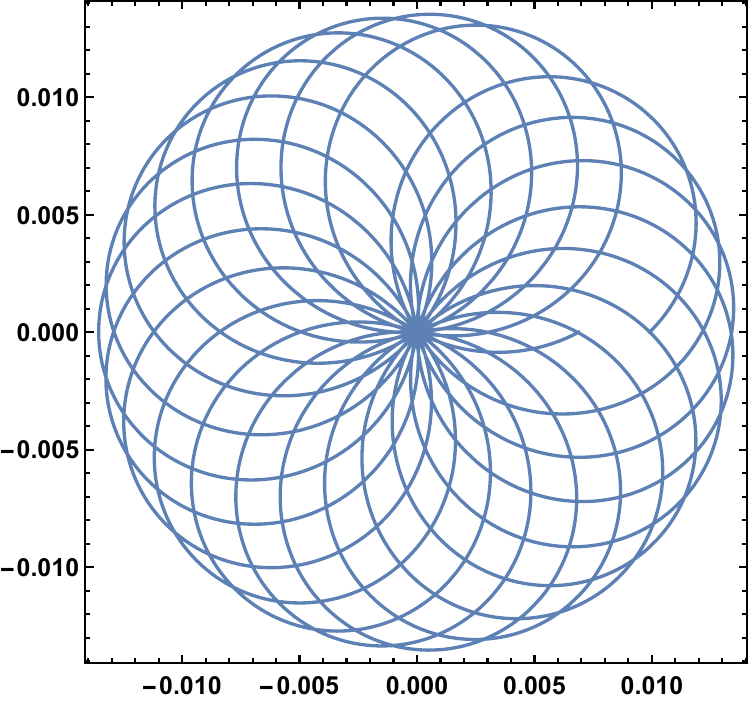}
\includegraphics[height=6cm,width=6cm]{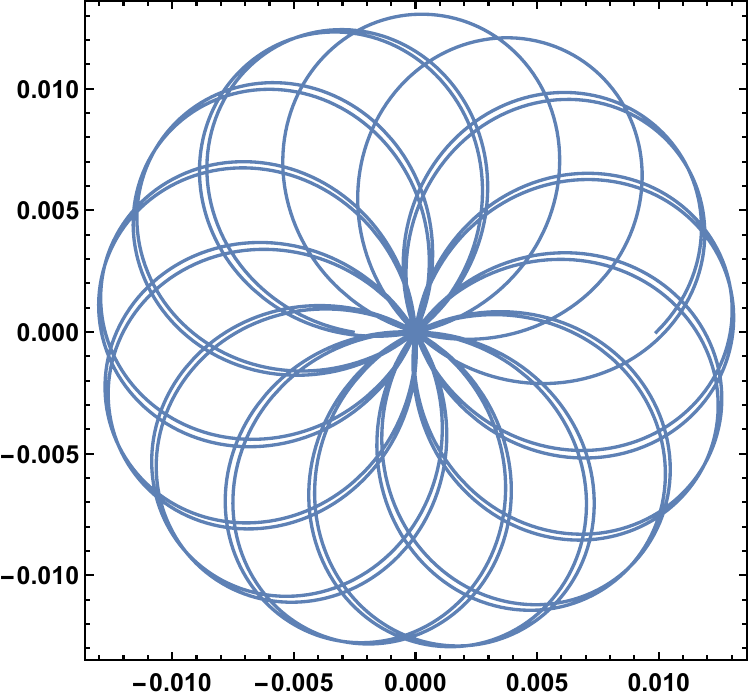}\\
\hspace{1cm}(a) $\alpha=r_s=\rho_s=0.2$\hspace{2cm} (b) $\alpha=0.4$ and $r_s=\rho_s=0.2$\\ \vspace{0.2cm}
\includegraphics[height=6cm,width=6cm]{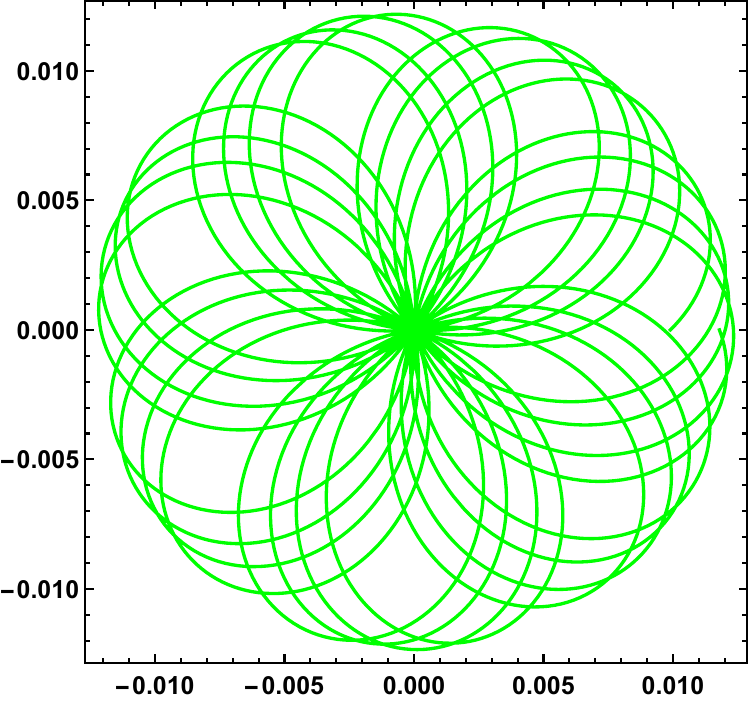}
\includegraphics[height=6cm,width=6cm]{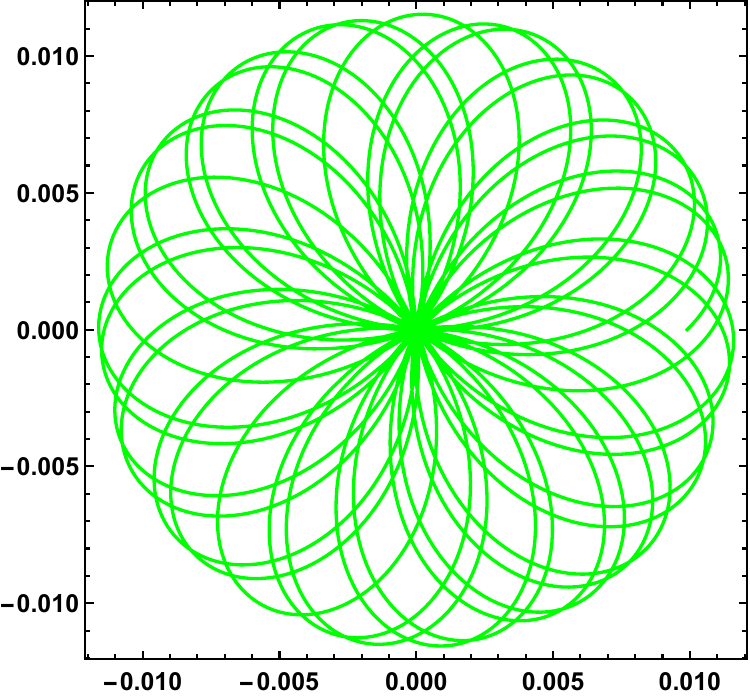}\\
\hspace{1cm}(c) $\alpha=r_s=0.4$ and $\rho_s=0.2$\hspace{2cm} (d) $\alpha=r_s=\rho_s=0.4$\\
\caption{{Light trajectories for different values of the model parameters with fixed mass $M=0.1$. The initial boundary conditions are chosen as $u(0)=0.01$ and $u'(0)=0.01$.}}
\label{fig9}
\end{figure}

Figure \ref{fig9} displays the photon trajectories for different choices of the parameters $(\alpha,r_s,\rho_s)$ with fixed BH mass $M=0.1$. The trajectories reveal the strong influence of the geometric deformation parameter and the Hernquist dark matter halo on the orbital structure of light rays propagating in the spacetime. In Fig. \ref{fig9}(a), for $\alpha=r_s=\rho_s=0.2$, the photon paths exhibit nearly symmetric bounded orbits with moderate angular deflection, indicating a balanced interplay between the Schwarzschild gravitational attraction and the exponential matter contribution. In Fig. \ref{fig9}(b), increasing the deformation parameter to $\alpha=0.4$ while keeping $r_s=\rho_s=0.2$ produces more pronounced distortions in the trajectories, enhancing the curvature of the photon paths and indicating stronger spacetime deformation effects on null geodesics. 
In Fig. \ref{fig9}(c), for $\alpha=r_s=0.4$ and $\rho_s=0.2$, the trajectories become more extended and display larger orbital structures, showing that the increase of the scale parameter $r_s$ amplifies the radial influence of the dark matter halo over photon propagation. Finally, in Fig. \ref{fig9}(d), for $\alpha=r_s=\rho_s=0.4$, exhibits the most intricate orbital configuration, with highly curved and densely packed trajectories around the central region. This behavior demonstrates that the combined increase of the halo density and geometric deformation intensifies gravitational lensing and strengthens the trapping of photons near the compact object.



\subsubsection{Particular cases}\label{S3-2-1}

The limiting configurations of the orbital equation provide useful physical insight into the separate roles played by the BH mass, the Hernquist halo parameters, and the constant deformation parameter in the propagation of photons. Starting from Eq. \eqref{orbit_u_new}, one notes that the motion of light is controlled by three distinct contributions, i.e., the Schwarzschild term proportional to $M$, the exponential factor associated with the dark matter distribution, and the constant shift governed by $\alpha$. Each of these terms modifies the optical geometry in a different way, affecting the bending angle, the possible existence of circular photon orbits, and the structure of photon trapping near the compact object.

First, let us consider the limit $\rho_s=0$. In this case, the dark matter density contribution is removed and the metric function term reduces to
\begin{align}
    f(r)=1-\frac{2M}{r}-\alpha .
\end{align}
This geometry corresponds to a Schwarzschild-like spacetime modified by a constant deformation parameter. Particularly, the parameter $\alpha$ changes the asymptotic value of the metric function, producing a global shift in the optical structure of the spacetime. For $\alpha=0$, the usual Schwarzschild solution is immediately recovered. The corresponding photon orbital equation reduces to
\begin{align}
    \left(\frac{du}{d\phi}\right)^{2}=\frac{1}{b^{2}}-(1-\alpha)u^{2}+2Mu^{3}.
\end{align}
This expression shows that $\alpha$ does not generate a Schwarzschild-like $u^{3}$ contribution. Instead, it modifies the coefficient of the quadratic term in $u$, changing the angular structure of the photon trajectory. Physically, the mass parameter $M$ remains responsible for the local strong-field attraction and for the nonlinear bending term $2Mu^{3}$, while $\alpha$ changes the effective angular propagation of the light ray. The corresponding circular photon orbit follows from the condition
\begin{align}
    r f'(r)-2f(r)=0,
\end{align}
which gives
\begin{align}
    r_{\rm ph}=\frac{3M}{1-\alpha}.
\end{align}
Therefore, for $0<\alpha<1$, the photon sphere is shifted to larger radii in comparison with the Schwarzschild value $r_{\rm ph}=3M$. This indicates that the constant deformation enlarges the optical trapping region around the black hole.

An second regime is obtained when $\alpha=0$. In this case, the metric function is
\begin{align}
    f(r)=\exp\left(-\frac{4\pi r_s^{3}\rho_s}{r+r_s}\right)
    -\frac{2M}{r}.
\end{align}
This limit isolates the effect of the Hernquist dark matter halo on the Schwarzschild geometry. Contrary to the case with nonzero $\alpha$, the spacetime is asymptotically flat, since the exponential factor approaches unity for $r\to\infty$. However, the approach to the asymptotic regime is modified by the halo parameters. Indeed, at large distances one has
\begin{align}
    \exp\left(-\frac{4\pi r_s^{3}\rho_s}{r+r_s}\right)
    \simeq 1-\frac{4\pi r_s^{3}\rho_s}{r}+\cdots,
\end{align}
so that the metric behaves as
\begin{align}
    f(r)\simeq 1-\frac{2M+4\pi r_s^{3}\rho_s}{r}+\cdots.
\end{align}
Thus, the Hernquist halo contributes to the effective gravitational mass felt by photons in the weak-field region. The orbital equation becomes
\begin{align}
    \left(\frac{du}{d\phi}\right)^{2}=\frac{1}{b^{2}}
    -u^{2}\exp\left(-\frac{4\pi r_s^{3}\rho_s u}{1+r_s u}
    \right)+2Mu^{3}.
\end{align}
In this case, the exponential term modifies the photon dynamics in a genuinely radial-dependent way. Unlike the constant parameter $\alpha$, the halo contribution changes with $r$, affecting both the weak-field deflection and the strong-field region near the photon sphere. Consequently, the variations of $r_s$ and $\rho_s$ may shift the circular photon orbit and modify the critical impact parameter associated with photon capture.

Another instructive case is obtained by taking $M=0$. The metric function then reads
\begin{align}
    f(r)=\exp\!\left(-\frac{4\pi r_s^{3}\rho_s}{r+r_s}\right)-\alpha.
\end{align}
In this configuration, the Schwarzschild contribution is absent and the photon motion is governed only by the dark matter distribution and by the constant deformation parameter. Thus, the orbital equation becomes
\begin{align}
    \left(\frac{du}{d\phi}\right)^{2}=\frac{1}{b^{2}}-u^{2}\exp\!\left(-\frac{4\pi r_s^{3}\rho_s u}{1+r_s u}\right)+\alpha u^{2}.
\end{align}
This limit is useful because it separates the effect of the environment from the usual BH mass contribution. The exponential term represents the radial gravitational influence of the Hernquist halo, while the $\alpha u^2$ term changes the global angular behavior of null rays. Therefore, even in the absence of the Schwarzschild mass, photon trajectories may still be deformed by the matter distribution and by the nontrivial asymptotic structure induced by $\alpha$. However, the physical interpretation differs from the Schwarzschild case. Thus, one notes that the bending is no longer produced by a central $1/r$ mass term, but by the extended halo profile and by the global geometric deformation of the spacetime.

Finally, let us consider the limit $r_s\to 0$ with finite $\rho_s$. Since the combination $4\pi r_s^3\rho_s$ tends to zero, the exponential correction becomes trivial,
\begin{align}
    \exp\!\left(-\frac{4\pi r_s^{3}\rho_s}{r+r_s}\right)
    \longrightarrow 1 .
\end{align}
Thus, the metric function reduces again to
\begin{align}
    f(r)=1-\frac{2M}{r}-\alpha ,
\end{align}
and the photon dynamics coincide with those of the Schwarzschild-like geometry modified by the constant deformation parameter. This result shows that, for finite $\rho_s$, the halo contribution disappears when the scale radius tends to zero. Hence, the parameter $r_s$ controls the spatial extension of the dark matter effect. Furthermore, when $r_s$ is sufficiently small, the exponential correction becomes negligible, while for larger $r_s$ the halo affects photon propagation over a wider radial region.

\section{Thermodynamics}\label{S4}

In this section, we investigate the thermodynamic properties of the black hole described by the metric function
\begin{align}
f(r)=\exp\!\left(-\frac{4\pi r_s^{3}\rho_s}{r+r_s}\right)-\frac{2M}{r}-\alpha.
\label{metric_thermo_new}
\end{align}
This geometry includes a density-dependent exponential core characterized by the parameters $(r_s,\rho_s)$, together with a constant deformation parameter $\alpha$, which shifts the effective potential.  
Such non-polynomial modifications naturally arise in models where matter fields or quantum-gravity-inspired corrections regulate the ultraviolet behavior of the spacetime geometry.  
Our goal is to determine the key thermodynamic quantities associated with this spacetime and to clarify how the exponential structure influences the thermal behavior and stability of the black hole.

\subsection{BH Mass and Horizon Condition}\label{S4-1}

The event horizon $r_+$ is defined as the largest real root of $f(r_+)=0$, which determines the causal boundary of the BH spacetime \cite{GR2,BT1,BT2}. Substituting Eq.~\eqref{metric_thermo_new}, the condition $f(r_+)=0$ becomes
\begin{align}
    \exp\!\left(-\frac{4\pi r_s^{3}\rho_s}{r_+ + r_s}\right)-\frac{2M}{r_+}-\alpha=0.
\label{horizon_equation_expanded}
\end{align}
The above relation incorporates the corrections induced by the Hernquist dark matter halo profile \cite{DH12,DH10,DH11,DH13}. 
Solving for $M$ yields the ADM mass, viz.,
\begin{align}
    M=\frac{r_+}{2}\left[\exp\!\left(-\frac{4\pi r_s^{3}\rho_s}{r_+ + r_s}\right)-\alpha\right].
\label{mass_general_new}
\end{align}
The mass therefore depends explicitly on the exponential density profile, providing a thermodynamic imprint of the underlying matter configuration and modifying the gravitational properties of the black hole relative to the standard Schwarzschild geometry.
\begin{figure}[ht!]
\begin{center}
\begin{tabular}{ccc}
\includegraphics[height=4cm,width=5cm]{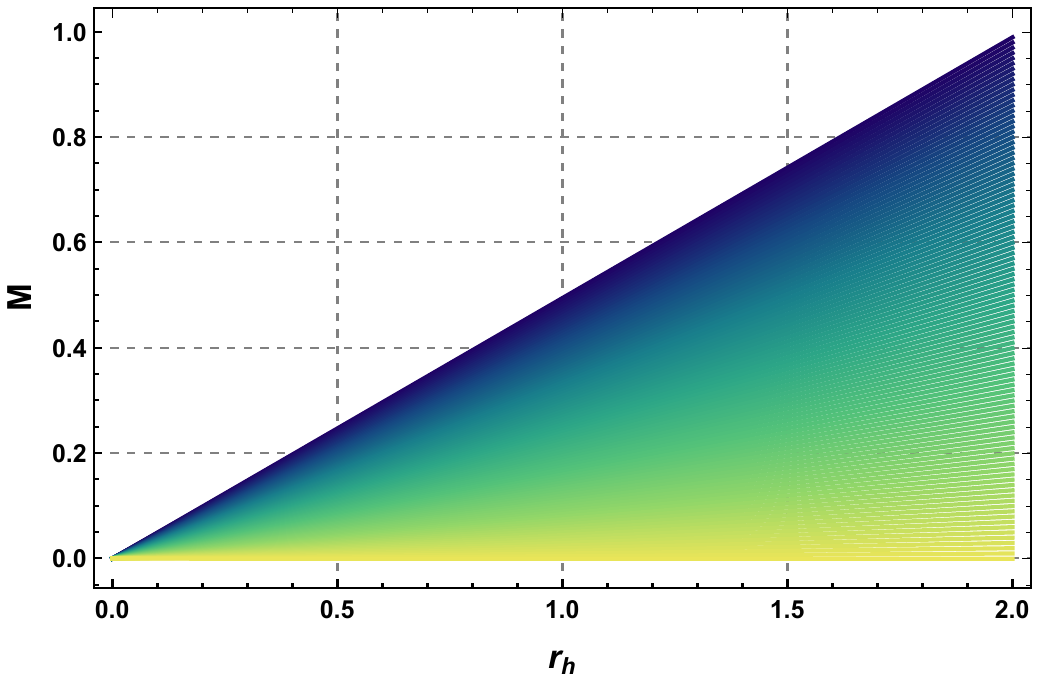} 
\includegraphics[height=4cm]{fig1aa.pdf}
\includegraphics[height=4cm,width=5cm]{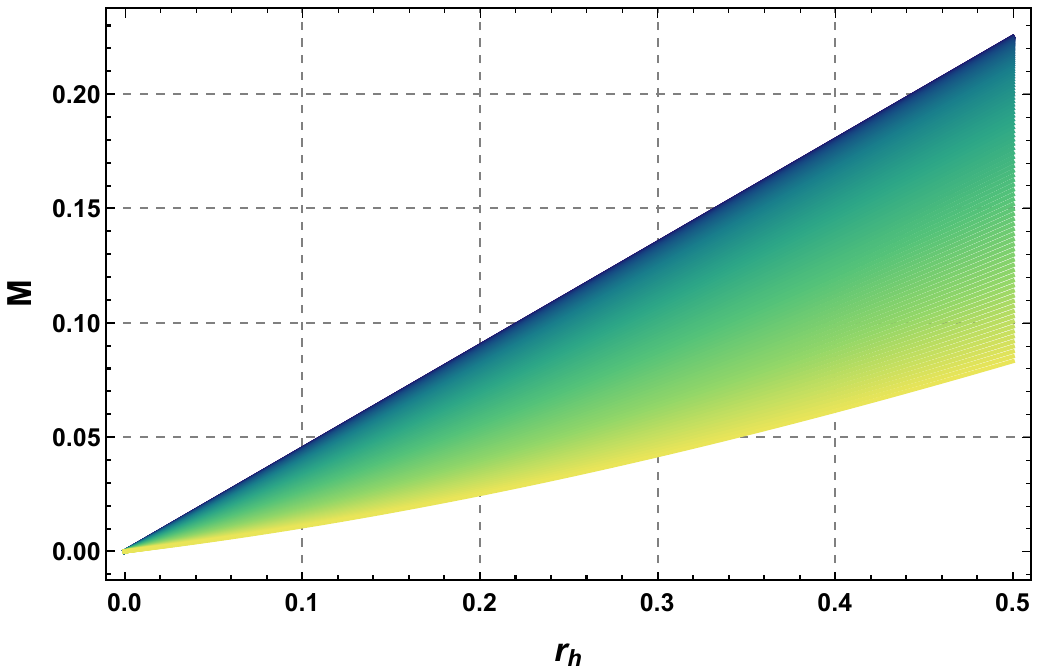} 
\includegraphics[height=4cm]{fig1bb.pdf}
\includegraphics[height=4cm,width=5cm]{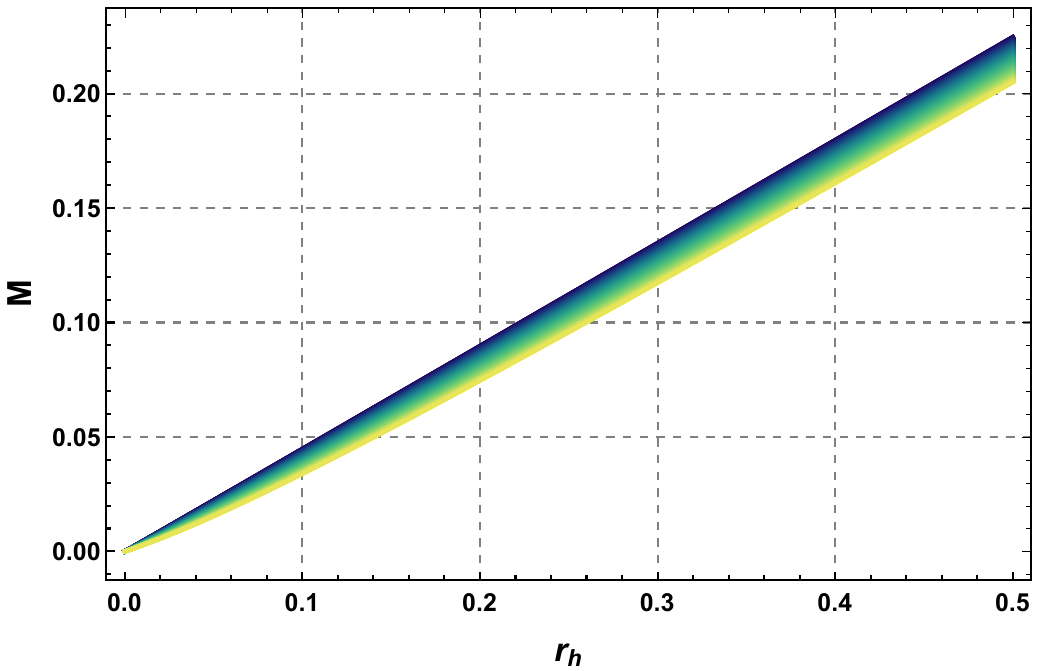} 
\includegraphics[height=4cm]{fig1cc.pdf}\\
\hspace{0.5cm} (a) $r_s=\rho_s=0.5$ \hspace{2.5cm} (b) $\alpha=\rho_s=0.5$ \hspace{2.5cm} (c) $r_s=\alpha=0.5$\\
\end{tabular}
\end{center}
\caption{The behavior of the ADM mass.\label{fig03}}
\end{figure}

Figure \ref{fig03} illustrates the behavior of the ADM mass $M$ as a function of the horizon radius $r_h$ for different values of the parameters $(\alpha,r_s,\rho_s)$ associated with the deformed black hole spacetime embedded in a Hernquist dark matter halo. In Fig. \ref{fig03}(a), with fixed values $r_s=\rho_s=0.5$, the increase of the deformation parameter $\alpha$ leads to a significant reduction in the ADM mass for a given horizon radius. This behavior follows directly from the negative contribution of $\alpha$ in the horizon equation and indicates that the geometric deformation weakens the effective mass contribution required to sustain the event horizon. Consequently, larger values of $\alpha$ generate lighter black hole configurations for the same horizon scale.

In Fig. \ref{fig03}(b), for fixed $\alpha=\rho_s=0.5$, variations of the scale parameter $r_s$ modify the slope of the mass function. As $r_s$ increases, the ADM mass decreases for fixed $r_h$, showing that the spatial extension of the Hernquist halo alters the gravitational energy distribution surrounding the black hole. This effect demonstrates that the scale radius controls how strongly the dark matter halo contributes to the global spacetime geometry. In Fig. \ref{fig03}(c) presents the influence of the density parameter $\rho_s$ for fixed $r_s=\alpha=0.5$. The figure shows that increasing $\rho_s$ slightly decreases the ADM mass, although the effect is weaker compared to the influence of $\alpha$ and $r_s$. Physically, this behavior reflects the exponential suppression induced by the dark matter density profile, which modifies the effective gravitational contribution near the horizon.

\subsection{Hawking Temperature}\label{S4-2}

The Hawking temperature is obtained from the surface gravity \cite{BT2,BT3,BT4}, i.e., 
\begin{align}
    T_H=\frac{f'(r_+)}{4\pi},
\label{temperature_def}
\end{align}
where $f'(r)$ is the radial derivative of the metric function. 
This relation establishes the connection between the geometrical structure of the event horizon and the thermal radiation emitted by the BH \cite{BT1,BT2,BT5,BT18}. From Eq.~\eqref{metric_thermo_new}, one obtains
\begin{align}
    f'(r)=\frac{2M}{r^2}-\exp\!\left(-\frac{4\pi r_s^{3}\rho_s}{r+r_s}\right)\left(\frac{4\pi r_s^{3}\rho_s}{(r+r_s)^2}\right).
\label{A_prime_general}
\end{align}
Evaluating at $r=r_+$ and using the mass relation \eqref{mass_general_new}, we conclude that
\begin{align}
    T_H=\frac{1}{4\pi}\left[\frac{1}{r_+}\left(\exp\!\left(-\frac{4\pi r_s^{3}\rho_s}{r_+ + r_s}\right)-\alpha\right)-\exp\!\left(-\frac{4\pi r_s^{3}\rho_s}{r_+ +r_s}\right)\frac{4\pi r_s^{3}\rho_s}{(r_+ + r_s)^2}\right].
\label{temperature_final}
\end{align}
This shows how both the deformation parameter $\alpha$ and the exponential core determine the thermal radiation.

\begin{figure}[ht!]
\begin{center}
\begin{tabular}{ccc}
\includegraphics[height=4cm,width=5cm]{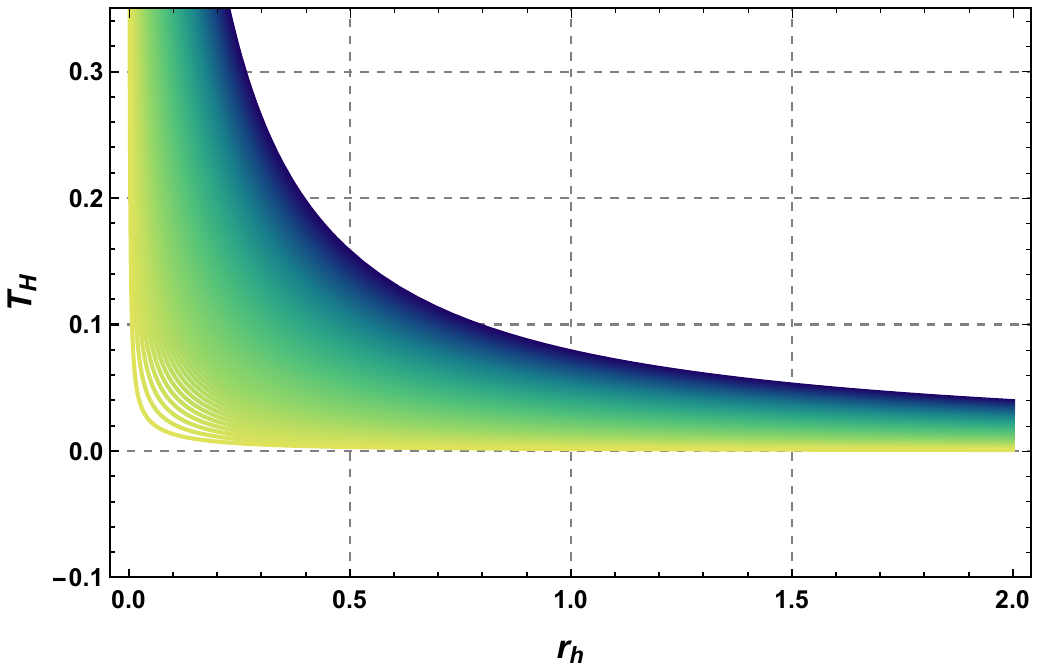} 
\includegraphics[height=4cm]{fig1aa.pdf}
\includegraphics[height=4cm,width=5cm]{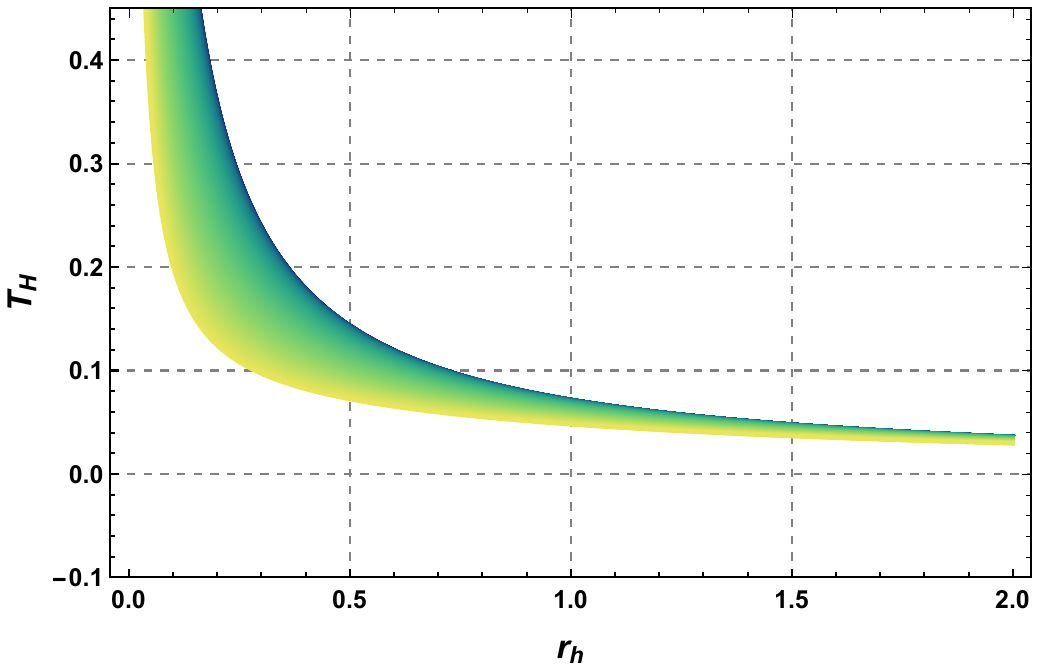} 
\includegraphics[height=4cm]{fig1bb.pdf}
\includegraphics[height=4cm,width=5cm]{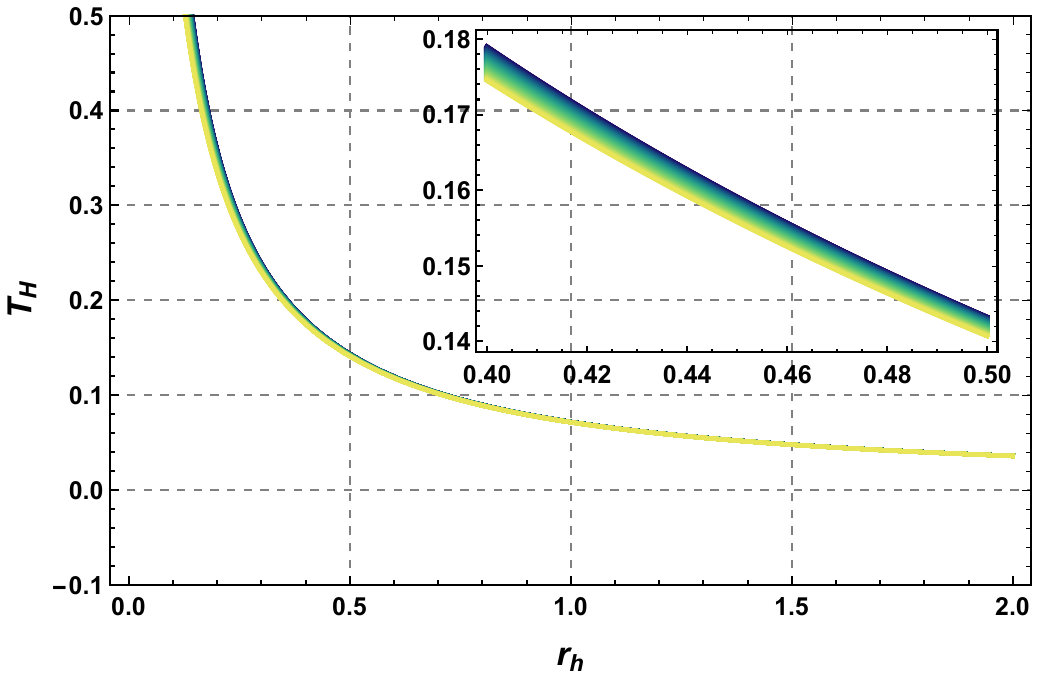} 
\includegraphics[height=4cm]{fig1cc.pdf}\\
\hspace{0.25cm} (a) $r_s=\rho_s=0.5$ \hspace{2.5cm} (b) $\alpha=\rho_s=0.5$ \hspace{2.5cm} (c) $r_s=\alpha=0.5$\\
\end{tabular}
\end{center}
\caption{The Hawking temperature $T_H$ vs. $r_H$.\label{fig04}}
\end{figure}

Figure~\ref{fig04} shows the behavior of the Hawking temperature $T_H$ as a function of the horizon radius $r_h$ for different values of the parameters $(\alpha,r_s,\rho_s)$ characterizing the deformed black hole surrounded by a Hernquist dark matter halo. In all figures, the Hawking temperature decreases monotonically as the horizon radius increases, which is consistent with the standard thermodynamic behavior of Schwarzschild-like black holes, where larger black holes radiate more slowly and are thermodynamically colder.

In Fig. \ref{fig04}(a), with fixed values $r_s=\rho_s=0.5$, increasing the deformation parameter $\alpha$ significantly suppresses the Hawking temperature, especially in the small-horizon regime. This behavior indicates that the geometric deformation weakens the surface gravity at the event horizon, thereby reducing the thermal radiation emitted by the BH. Consequently, larger values of $\alpha$ tend to generate colder and more thermodynamically stable configurations.
Fig. \ref{fig04}(b), corresponding to fixed $\alpha=\rho_s=0.5$, illustrates the influence of the scale parameter $r_s$. The results show that increasing $r_s$ slightly lowers the temperature profile, particularly near the strong-field region. Physically, this reflects the effect of the extended dark matter halo, whose radial distribution modifies the near-horizon geometry and reduces the effective gravitational gradient responsible for Hawking radiation.
In Fig. \ref{fig04}(c), for fixed $r_s=\alpha=0.5$, the density parameter $\rho_s$ also affects the thermal behavior of the BH. As shown in the plots, increasing $\rho_s$ produces a small but noticeable reduction in the Hawking temperature. This effect arises from the exponential density correction in the metric function, which smooths the spacetime geometry near the horizon and decreases the surface gravity.

\subsection{Entropy}\label{S4-3}

Since the geometry is static and spherically symmetric, the Bekenstein-Hawking entropy is given by the standard area law~\cite{BT1,BT2,BT5}, i.e., 
\begin{align}
    S=\pi r_+^2.
\label{entropy}
\end{align}
This expression establishes that the entropy of the BH is proportional to the area of the event horizon, representing one of the fundamental results of semiclassical gravity and BH thermodynamics \cite{BT1,BT2,BT5}. Although the exponential modification alters the mass and temperature, it does not modify the universal area-entropy relation, reflecting the fact that $f(r)$ remains regular and well-defined at the horizon.

\subsection{Gibbs Free Energy}\label{S4-4}

taking into account the entropy \eqref{entropy}, we note that the Gibbs free energy in the canonical ensemble is
\begin{align}
G = M - T_H S.
\label{gibbs_def}
\end{align}

By using Eqs.~\eqref{mass_general_new}, \eqref{temperature_final}, and \eqref{entropy}, we obtain
\begin{align}
G(r_+)=\frac{r_+}{2}\left[\exp\!\left(-\frac{4\pi r_s^{3}\rho_s}{r_+ + r_s}\right)-\alpha\right]-\pi r_+^2 T_H.
\label{gibbs_final}
\end{align}
The resulting free-energy profile determines whether the system undergoes Hawking-Page-like transitions when analyzed against an appropriate thermodynamic background.
\begin{figure}[ht!]
\begin{center}
\begin{tabular}{ccc}
\includegraphics[height=4cm,width=5cm]{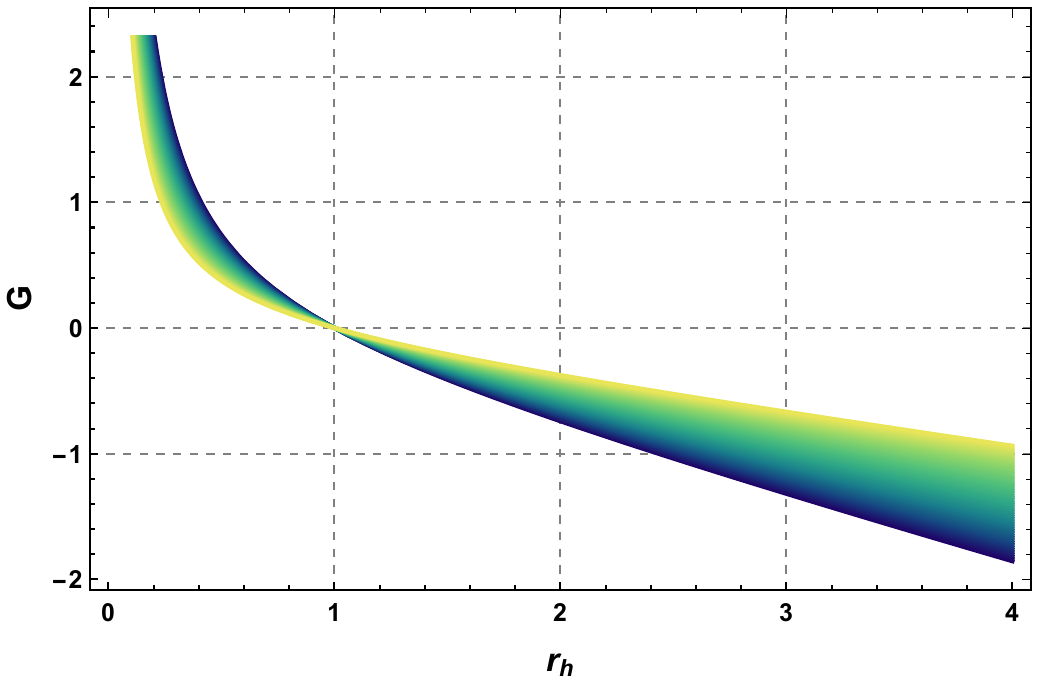} 
\includegraphics[height=4cm]{fig1aa.pdf}
\includegraphics[height=4cm,width=5cm]{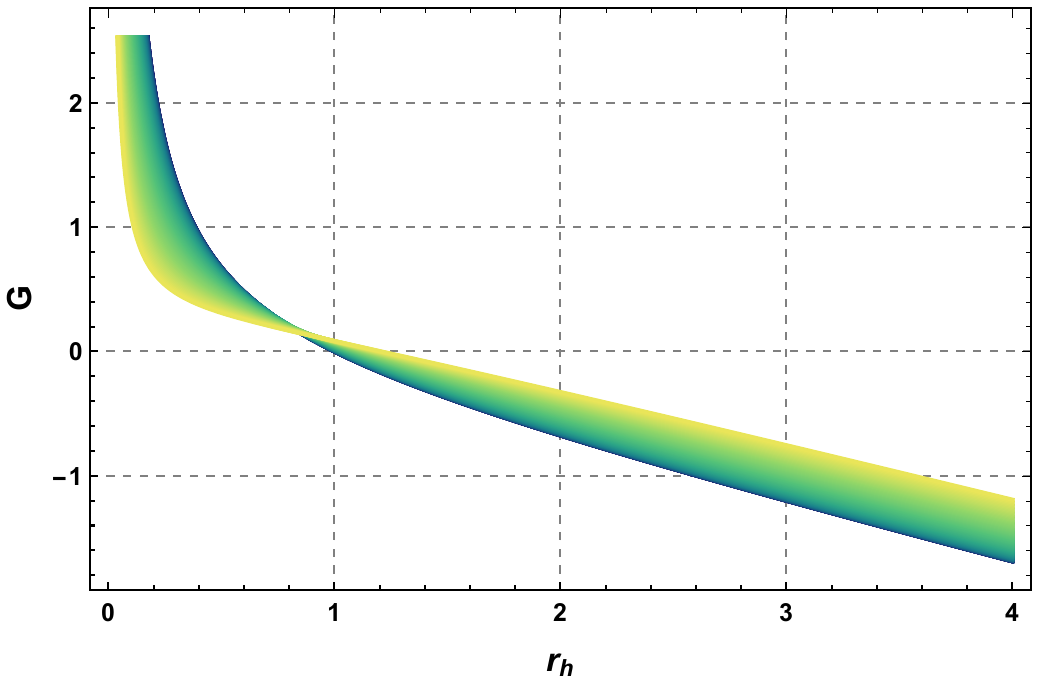} 
\includegraphics[height=4cm]{fig1bb.pdf}
\includegraphics[height=4cm,width=5cm]{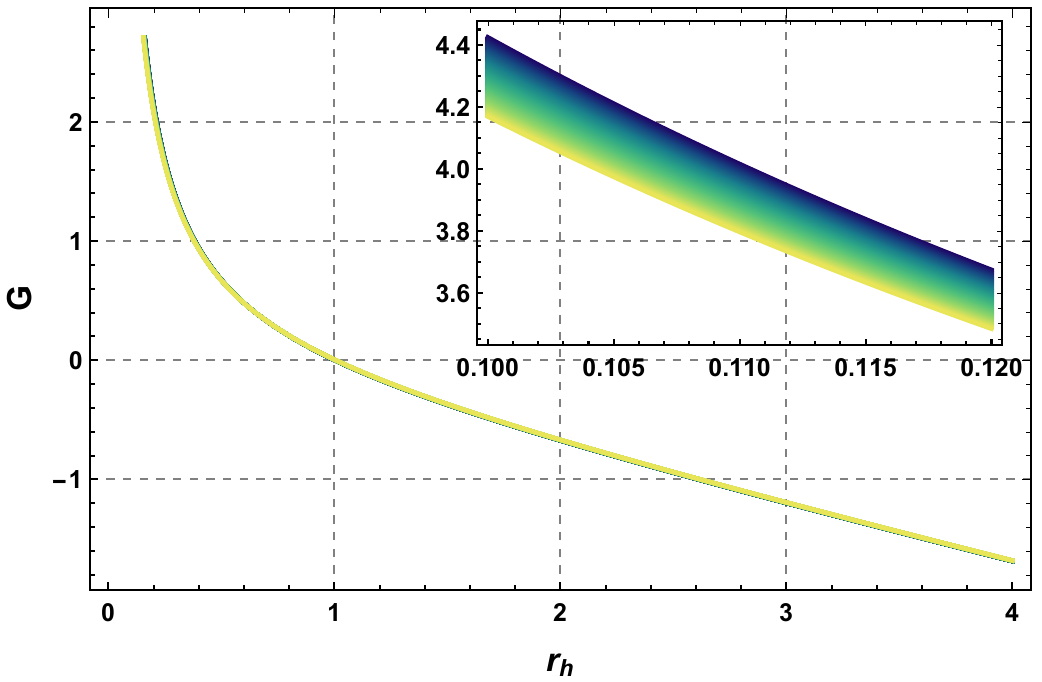} 
\includegraphics[height=4cm]{fig1cc.pdf}\\
\hspace{-0.2cm} (a) $r_s=\rho_s=0.5$ \hspace{2.5cm} (b) $\alpha=\rho_s=0.5$ \hspace{2.5cm} (c) $r_s=\alpha=0.5$\\
\end{tabular}
\end{center}
\caption{The Gibbs free energy vs. $r_h$.\label{fig06}}
\end{figure}

Figure~\ref{fig06} presents the behavior of the Gibbs free energy $G$ as a function of the horizon radius $r_h$ for different values of the parameters $(\alpha,r_s,\rho_s)$ associated with the deformed black hole spacetime immersed in a Hernquist dark matter halo. In all Fig.s, the Gibbs free energy decreases monotonically with increasing horizon radius, changing from positive to negative values. This transition indicates the existence of a thermodynamic threshold separating less stable small black holes from thermodynamically favored large black hole configurations. The region where $G<0$ corresponds to globally stable states, suggesting the possibility of Hawking--Page-type phase transitions in the system. In Fig. \ref{fig06}(a), with fixed values $r_s=\rho_s=0.5$, increasing the deformation parameter $\alpha$ shifts the Gibbs free energy toward lower values for large horizon radii. Physically, this behavior shows that the geometric deformation enhances the thermodynamic stability of the black hole by favoring states with lower free energy. The effect becomes more pronounced in the large-horizon regime, where the contribution of $\alpha$ dominates the asymptotic thermodynamic structure. In Fig. \ref{fig06}(b), corresponding to fixed $\alpha=\rho_s=0.5$, illustrates the influence of the scale parameter $r_s$. As $r_s$ increases, the free-energy curves are slightly shifted, modifying the point at which the Gibbs free energy changes sign. This result indicates that the spatial extension of the dark matter halo directly affects the phase structure of the black hole and alters the balance between thermal radiation and gravitational energy. In Fig. \ref{fig06}(c), for fixed $r_s=\alpha=0.5$, the density parameter $\rho_s$ produces small corrections to the Gibbs free energy, as emphasized in the inset plot. Increasing $\rho_s$ slightly lowers the free energy, indicating that denser dark matter halos contribute to more thermodynamically favorable black hole configurations.

\subsection{Heat Capacity and Thermal Stability}\label{S4-5}

The heat capacity at constant mass is
\begin{align}
    C=\frac{dM}{dT_H}=\frac{dM/dr_+}{dT_H/dr_+}.
\label{heat_capacity_def}
\end{align}
The derivative of the mass function is
\begin{align}
    \frac{dM}{dr_+}=\frac{1}{2}\left[\exp\!\left(-\frac{4\pi r_s^{3}\rho_s}{r_+ + r_s}\right)-\alpha\right]+\frac{r_+}{2}\exp\!\left(-\frac{4\pi r_s^{3}\rho_s}{r_+ +r_s}\right)\left(\frac{4\pi r_s^{3}\rho_s}{(r_+ + r_s)^2}\right).
\label{dM_dr}
\end{align}

The derivative of the temperature is obtained by differentiating Eq.~\eqref{temperature_final} with respect to $r_+$, which leads us to  
\begin{align}
    \frac{dT_H}{dr_+}=\frac{\alpha}{4\pi r_+^{2}}+e^{-\frac{4\pi r_s^{3}\rho_s}{r_++r_s}}\left(-\frac{1}{4\pi r_+^{2}}-\frac{2 r_s^{3}\rho_s}{(r_++r_s)^{3}}\right)+\frac{4\pi r_s^{3}\rho_s\,e^{-\frac{4\pi r_s^{3}\rho_s}{r_++r_s}}\left(\frac{1}{4\pi r_+}+\frac{r_s^{3}\rho_s}{(r_++r_s)^{2}}
\right)}{(r_++r_s)^{2}},
\label{dT_dr}
\end{align}
where $\mathcal{F}$ is a fully analytic but lengthy function involving exponential factors and rational powers of $(r_+ + r_s)$.  

The heat capacity then reads
\begin{align}
C=\frac{2 \pi^{2} r_+^{3} (r_++r_s)^{4}\left[-\frac{\alpha}{4\pi r_+}+e^{-\frac{4\pi r_s^{3}\rho_s}{r_++r_s}}\left(\frac{1}{4\pi r_+}+\frac{r_s^{3}\rho_s}{(r_++r_s)^{2}}
\right)\right]}{(r_++r_s)^{4}\alpha+4\pi r_+ r_s^{3}\rho_s\,e^{-\frac{4\pi r_s^{3}\rho_s}{r_++r_s}}\left[(r_++r_s)^{2} + 4\pi r_+ r_s^{3}\rho_s\right]-e^{-\frac{4\pi r_s^{3}\rho_s}{r_++r_s}}(r_++r_s)\left[(r_++r_s)^{3} + 8\pi r_+^{2} r_s^{3}\rho_s\right]}.
\label{C_final_expression}
\end{align}

A divergence in $C$ signals a second-order phase transition, identifying the points at which the BH switches between stable ($C>0$) and unstable ($C<0$) configurations. In this model, the exponential density profile tends to regularize ultraviolet divergences and can shift or remove unstable branches, thereby altering the thermal phase structure relative to standard Schwarzschild-like geometries.

\begin{figure}[ht!]
\begin{center}
\begin{tabular}{ccc}
\includegraphics[height=4cm,width=4.5cm]{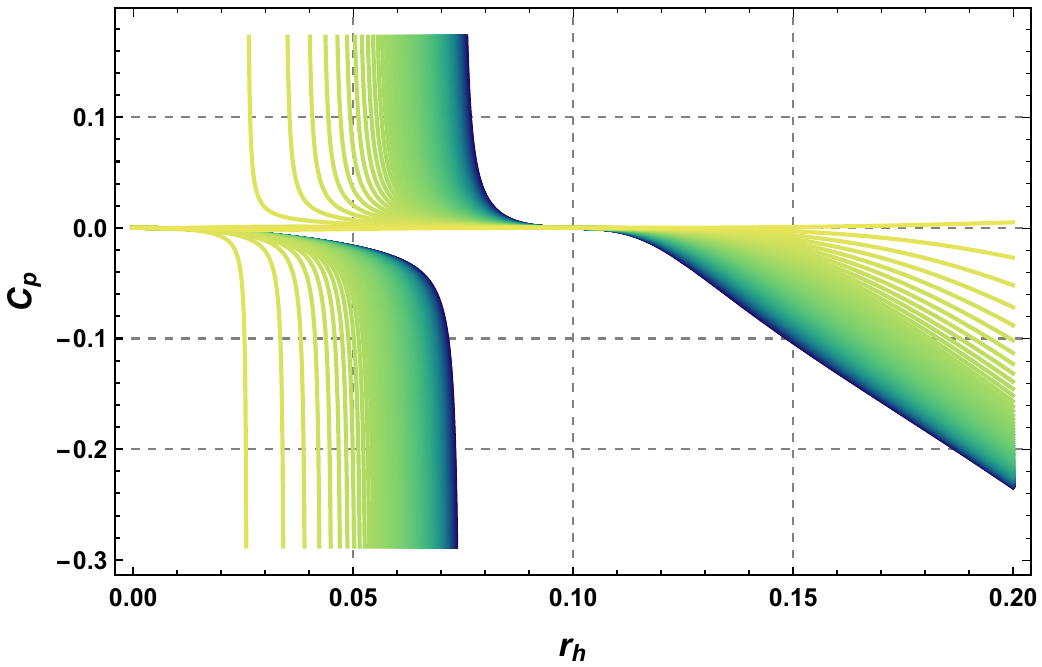} 
\includegraphics[height=4cm]{fig1aa.pdf}
\includegraphics[height=4cm,width=4.5cm]{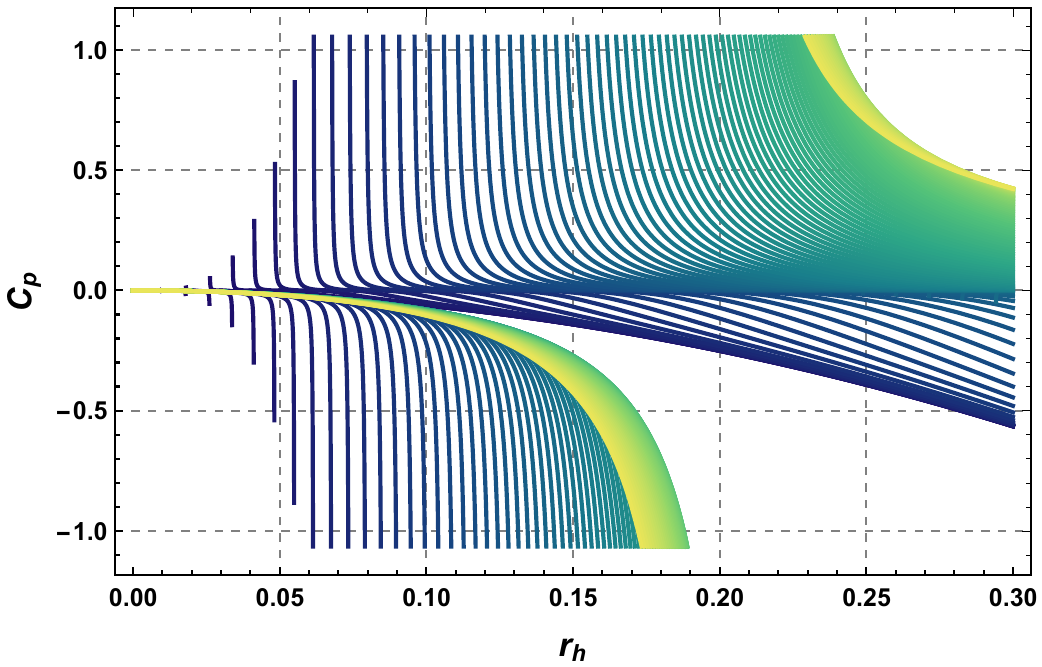} 
\includegraphics[height=4cm]{fig1bb.pdf}
\includegraphics[height=4cm,width=4.5cm]{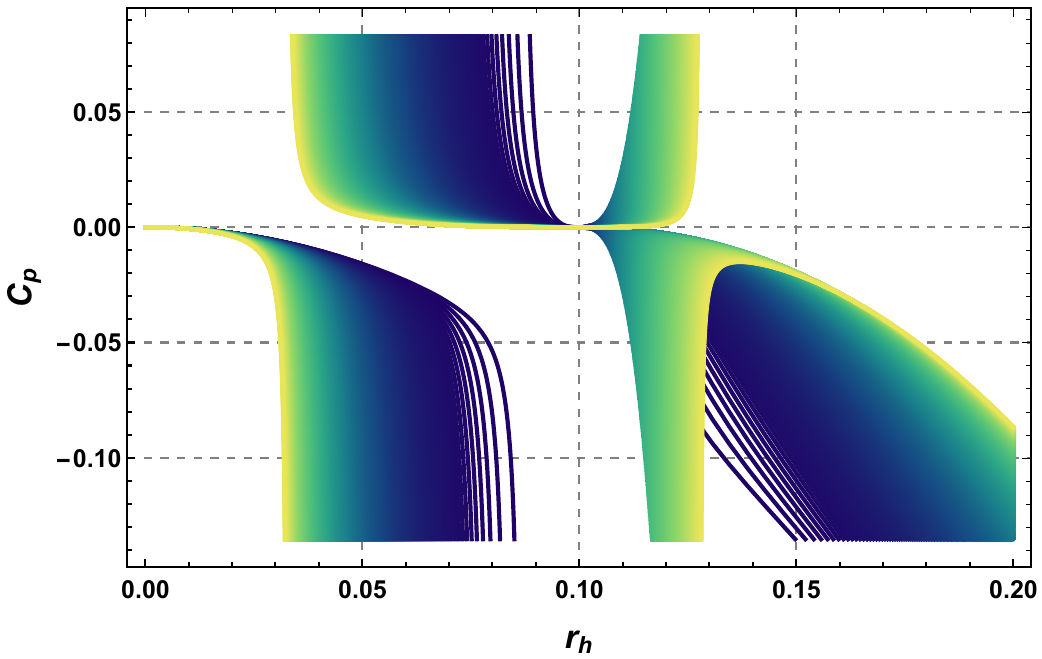} 
\includegraphics[height=4cm]{fig1cc.pdf}
\hspace{-2cm} (a) $r_s=\rho_s=0.5$ \hspace{2.5cm} (b) $\alpha=\rho_s=0.5$ \hspace{2.5cm} (c) $r_s=\alpha=0.5$\\
\end{tabular}
\end{center}
\caption{The solution of the heat capacity.\label{fig05}}
\end{figure}

Figure~\ref{fig05} illustrates the behavior of the heat capacity $C$ as a function of the event horizon radius $r_h$ for different values of the parameters $(\alpha,r_s,\rho_s)$. In all Fig.s, the heat capacity exhibits discontinuities and sign changes, which are characteristic signatures of second-order thermodynamic phase transitions. The divergence points of $C$ correspond to the roots of $dT_H/dr_h=0$, separating unstable configurations with negative heat capacity $(C<0)$ from locally stable configurations with positive heat capacity $(C>0)$. In Fig. \ref{fig05}(a), increasing the deformation parameter $\alpha$ significantly shifts the divergence points toward larger horizon radii and modifies the extent of the stable branches, indicating that the geometric correction strongly affects the thermal equilibrium structure of the BH. Fig. \ref{fig05}(b) shows that variations in the scale parameter $r_s$ produce a substantial deformation of the thermodynamic profile, enlarging the positive-$C$ region and therefore favoring thermodynamically stable configurations over a broader range of horizon radii. Meanwhile, Fig. \ref{fig05}(c) demonstrates that increasing the density parameter $\rho_s$ smooths the thermodynamic behavior and suppresses large negative divergences, suggesting that the Hernquist dark matter distribution acts as a regularizing contribution to the thermal dynamics. The combined effects of $(\alpha,r_s,\rho_s)$ generate a rich phase structure that departs significantly from the standard Schwarzschild case, showing that the surrounding dark matter halo and the geometric deformation parameter can stabilize the black hole against thermal fluctuations in specific parameter regimes.

\section{Scalar Perturbations}\label{S5}

Finally, let us investigate the evolution of a massless scalar field $\Phi$ propagating in the background geometry determined by the metric function (\ref{aa2}). The scalar perturbations provide a powerful means to probe the stability and dynamical response of black hole spacetimes, and they often serve as a first step toward computing quasinormal modes and understanding wave propagation in modified gravitational environments. The exponential term encodes the influence of the matter distribution characterized by $(r_s,\rho_s)$, the Schwarzschild term represents the mass contribution, and the deformation parameter $\alpha$ introduces an additional geometric shift which modifies both near- and far-field behavior.

The dynamics of a massless scalar field are governed by the covariant Klein-Gordon equation,
\begin{align}
    \frac{1}{\sqrt{-g}}\partial_\mu\!\left(\sqrt{-g}\,g^{\mu\nu}\partial_\nu\Phi\right)=0,
\label{KG_new}
\end{align}
where $g_{\mu\nu}$ is the spacetime metric and $g=\det(g_{\mu\nu})$.

To separate variables, we adopt the standard ansatz
\begin{align}
    \Phi(t,r,\theta,\phi)=e^{-i\omega t}Y_{\ell m}(\theta,\phi)\frac{\psi(r)}{r},
\label{ansatz_new}
\end{align}
with $Y_{\ell m}$ the spherical harmonics and $\omega$ the (complex) mode frequency. Substituting Eq.~(\ref{ansatz_new}) into Eq.~(\ref{KG_new}) and using the metric (\ref{aa2}), one obtains a Schrödinger-like wave equation,
\begin{align}
    \frac{d^{2}\psi(r_*)}{dr_*^{2}}+\left(\omega^{2}-\mathcal{V}(r)\right)\psi(r_*)=0,
\label{wave_new}
\end{align}
after introducing the tortoise coordinate,
\begin{align}
    r_*=\int\frac{dr}{f(r)}.
\label{tortoise_new}
\end{align}

The effective potential for scalar perturbations takes the general form
\begin{align}
    \mathcal{V}_{\text{scalar}}(r)=f(r)\left[\frac{\ell(\ell+1)}{r^{2}}+\frac{f'(r)}{r}\right],
\label{Vscalar_general_new}
\end{align}
where the prime denotes differentiation with respect to $r$.

Using the explicit form of the metric function (\ref{aa2}), the effective potential becomes
\begin{align}
    \mathcal{V}_{\text{scalar}}(r)&=\left[\exp\!\left(-\frac{4\pi r_s^{3}\rho_s}{r+r_s}\right)-\frac{2M}{r}-\alpha\right]\left[\frac{\ell(\ell+1)}{r^{2}}+\frac{1}{r}\frac{d}{dr}\left(\exp\!\left(-\frac{4\pi r_s^{3}\rho_s}{r+r_s}\right)-\frac{2M}{r}-\alpha\right)\right].
\label{Vscalar_explicit_new}
\end{align}

The structure of the potential is dictated by the interplay among the mass term $M$, the matter distribution parameters $(r_s,\rho_s)$ embedded within the exponential factor, and the geometric deformation parameter $\alpha$. The exponential profile modifies the near-horizon curvature and may shift the peak of the potential, whereas the parameter $\alpha$ induces a global shift in the asymptotic behavior. Together, these features govern the propagation of scalar waves, influence the quasinormal mode spectrum, and determine the stability properties of the spacetime under scalar perturbations.

\begin{figure}[ht!]
\begin{center}
\begin{tabular}{ccc}
\includegraphics[height=4cm,width=4.5cm]{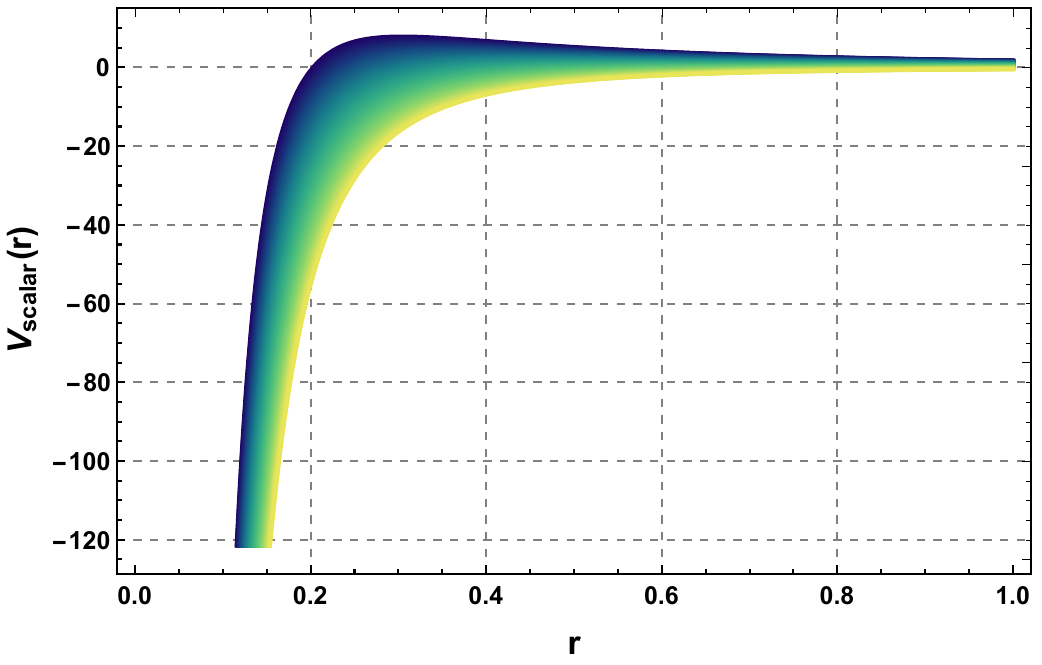} 
\includegraphics[height=4cm]{fig1aa.pdf}
\includegraphics[height=4cm,width=4.5cm]{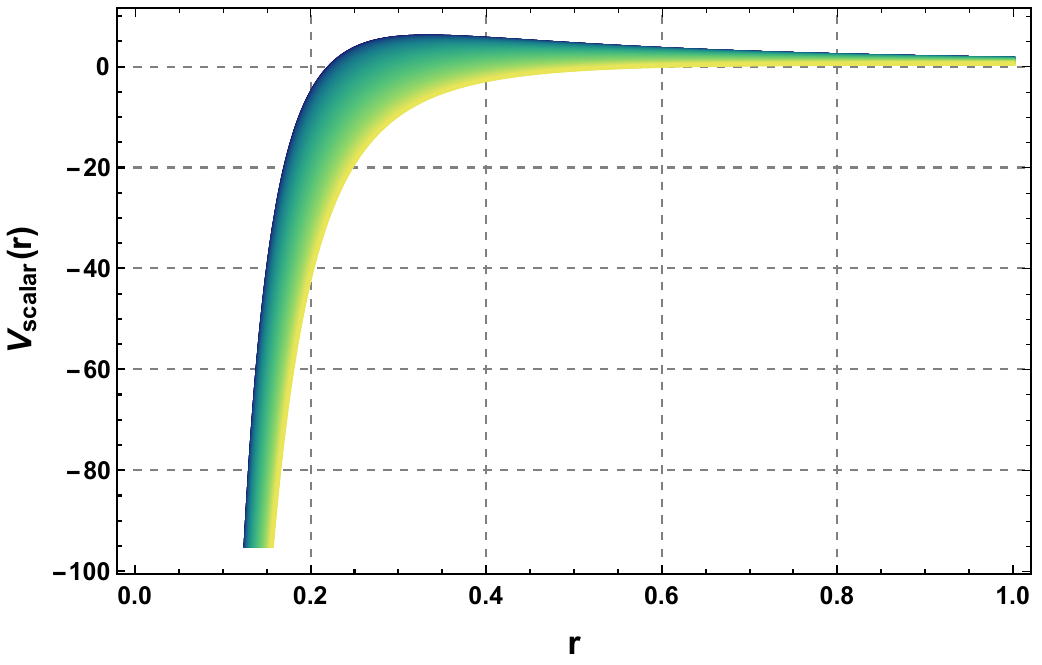} 
\includegraphics[height=4cm]{fig1bb.pdf}
\includegraphics[height=4cm,width=4.5cm]{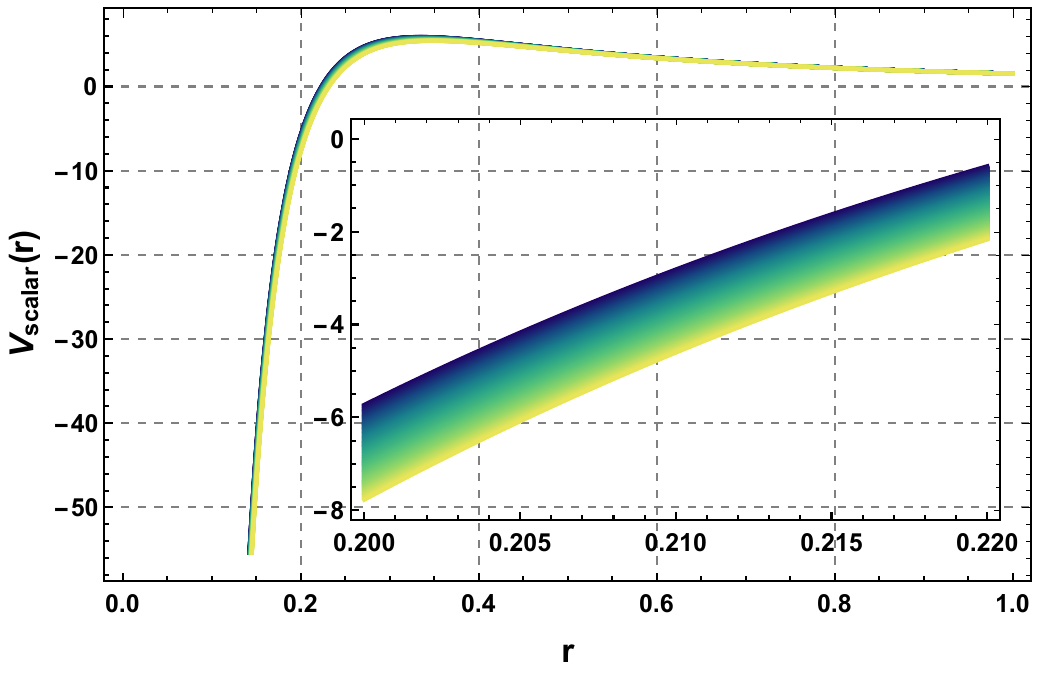} 
\includegraphics[height=4cm]{fig1cc.pdf}\\
\hspace{-0.15cm} (a) $r_s=\rho_s=0.5$ \hspace{2.5cm} (b) $\alpha=\rho_s=0.5$ \hspace{2.5cm} (c) $r_s=\alpha=0.5$\\
\end{tabular}
\end{center}
\caption{The behavior of the effective potential for scalar perturbations ($M=0.1$).\label{fig07}}
\end{figure}

Figure \ref{fig07} displays the behavior of the effective potential for scalar perturbations, $V_{\text{scalar}}(r)$, as a function of the radial coordinate for different values of the parameters $(\alpha,r_s,\rho_s)$. The profiles reveal the existence of a potential barrier outside the event horizon, which is a typical feature governing the propagation and scattering of scalar waves in black hole spacetimes. In Fig. \ref{fig07}(a), the increase of the deformation parameter $\alpha$ modifies both the height and width of the effective potential barrier, shifting the peak toward larger radial distances and altering the trapping region of scalar perturbations. This behavior indicates that the geometric deformation changes the effective curvature experienced by the scalar field and consequently affects the quasinormal spectrum and wave scattering properties. In Fig. \ref{fig07}(b), variations of the scale parameter $r_s$ significantly influence the asymptotic decay and localization of the potential barrier, showing that the characteristic scale of the Hernquist dark matter halo controls the spatial extension of the perturbative interaction region. Fig. \ref{fig07}(c) demonstrates that increasing the density parameter $\rho_s$ enhances the overall magnitude of the potential and smooths the near-horizon structure, as emphasized in the inset plot. This suggests that the dark matter distribution contributes to a partial regularization of the scalar effective potential and modifies the stability properties of the spacetime under scalar perturbations. 

\subsection{Particular cases}\label{S5-1}

To clarify the physical content of the scalar effective potential and the wave dynamics governed by Eq.~(\ref{wave_new}), it is useful to examine some limiting regimes of the metric function~(\ref{aa2}). These limits provide consistency checks of the general result and allow us to identify how the BH mass, the Hernquist dark matter distribution, and the constant deformation parameter $\alpha$ separately affect the propagation of scalar perturbations. Since the effective potential has the general structure
\begin{align}
    \mathcal{V}_{\text{scalar}}(r)=f(r)\left[\frac{\ell(\ell+1)}{r^{2}}+\frac{f'(r)}{r}\right],
\end{align}
any modification of the metric function changes the scalar dynamics in two different ways, i.e., through the multiplicative redshift factor $f(r)$ and through the derivative term $f'(r)$, which encodes the local radial variation of the gravitational field. Therefore, the height, width, and radial position of the potential barrier are directly controlled by the interplay among $M$, $(r_s,\rho_s)$, and $\alpha$.

In the vacuum limit, obtained by taking $\rho_s\rightarrow 0$, the contribution of the Hernquist matter distribution is removed. In this case, the exponential factor tends to unity and the metric function becomes
\begin{align}
    f(r)\Big|_{\rho_s=0}=1-\frac{2M}{r}-\alpha .
\end{align}
Consequently, the scalar effective potential reduces to
\begin{align}
    \mathcal{V}_{\text{scalar}}(r)\Big|_{\rho_s=0}=\left(1-\frac{2M}{r}-\alpha\right)\left[\frac{\ell(\ell+1)}{r^{2}}+\frac{2M}{r^{3}}\right].
\end{align}
This expression corresponds to a Schwarzschild-like geometry modified by the constant shift $\alpha$. For $\alpha=0$, the usual Schwarzschild result is recovered. When $\alpha\neq 0$, however, the asymptotic value of the metric function is no longer unity, but rather $1-\alpha$. Thus, the parameter $\alpha$ changes the global normalization of the geometry and modifies the centrifugal part of the scalar potential. Particularly, at large distances the leading contribution behaves as
\begin{align}
    \mathcal{V}_{\mathrm{scalar}}(r)\Big|_{\rho_s=0}\sim (1-\alpha)\frac{\ell(\ell+1)}{r^2}.
\end{align}
Hence, $\alpha$ does not introduce a new radial matter profile, but it changes the asymptotic propagation of scalar waves by rescaling the angular barrier. Physically, this means that the deformation parameter alters the effective scattering environment even in the absence of the dark matter halo.

Another relevant regime is the no-deformation limit, $\alpha\rightarrow0$. In this case, the metric function becomes
\begin{align}
f(r)\Big|_{\alpha=0}=\exp\!\left(-\frac{4\pi r_s^{3}\rho_s}{r+r_s}\right)-\frac{2M}{r}.
\end{align}
The corresponding scalar effective potential is
\begin{align}
    \mathcal{V}_{\text{scalar}}(r)\Big|_{\alpha=0}=\left[\exp\!\left(-\frac{4\pi r_s^{3}\rho_s}{r+r_s}\right)-\frac{2M}{r}\right]\left[\frac{\ell(\ell+1)}{r^{2}}+\frac{2M}{r^{3}}+\frac{4\pi r_s^{3}\rho_s}{r(r+r_s)^2}\exp\!\left(-\frac{4\pi r_s^{3}\rho_s}{r+r_s}\right)\right].
\end{align}
This limit isolates the effect of the Hernquist halo on scalar perturbations. Differently from the constant parameter $\alpha$, the dark matter contribution depends explicitly on the radial coordinate. Therefore, it does not merely shift the potential, but changes its shape. The exponential profile modifies the radial gradient of the metric function and may displace the maximum of the scalar potential barrier. Since this maximum controls the dominant scattering region and is closely related to the quasinormal-mode spectrum, the parameters $r_s$ and $\rho_s$ can modify both the oscillation frequencies and the damping rates of scalar perturbations. Larger values of $r_s$ extend the radial domain over which the halo affects wave propagation, whereas larger $\rho_s$ strengthens the exponential correction associated with the matter density.

In the large-radius regime, $r\gg r_s$, the exponential term can be expanded as
\begin{align}
    \exp\!\left(-\frac{4\pi r_s^{3}\rho_s}{r+r_s}\right) \simeq 1-\frac{4\pi r_s^{3}\rho_s}{r} +\mathcal{O}\!\left(\frac{1}{r^2}\right).
\end{align}
Thus, the metric function assumes the asymptotic form
\begin{align}
f(r)\simeq 1-\alpha-\frac{2M_{\rm eff}}{r}
+\mathcal{O}\!\left(\frac{1}{r^2}\right) \qquad \mathrm{with} \qquad M_{\rm eff}=M+2\pi r_s^{3}\rho_s .
\end{align}
This result shows that, from the viewpoint of a distant scalar wave, the Hernquist halo contributes as an effective mass correction. The scalar potential then behaves asymptotically as
\begin{align}
    \mathcal{V}_{\text{scalar}}(r) \simeq (1-\alpha)\frac{\ell(\ell+1)}{r^{2}}+\frac{2M_{\rm eff}\left[1-\alpha-\ell(\ell+1)\right]}{r^{3}}+\mathcal{O}\!\left(\frac{1}{r^4}\right).
\end{align}
The leading term is controlled by the angular momentum number $\ell$ and by the asymptotic deformation factor $1-\alpha$, whereas the subleading term contains the effective mass generated by both the central BH and the surrounding halo. Therefore, at large distances the scalar field does not distinguish the BH mass and the halo contribution independently at leading order in $1/r$; instead, it responds to the combined quantity $M_{\rm eff}$. This limit makes explicit how the dark matter halo modifies weak-field scattering, while $\alpha$ changes the asymptotic normalization of the potential barrier.

The near-core regime, $r\ll r_s$, must be interpreted with some care. In this limit, the exponential factor becomes approximately constant,
\begin{align}
    \exp\!\left(-\frac{4\pi r_s^{3}\rho_s}{r+r_s}\right)\simeq \exp\!\left(-4\pi r_s^{2}\rho_s\right),
\end{align}
and the metric function approaches
\begin{align}
    f(r)\simeq \exp\!\left(-4\pi r_s^{2}\rho_s\right)-\alpha-\frac{2M}{r}.
\end{align}
The Hernquist contribution is finite in this region and acts as a constant correction to the metric function. However, for $M\neq0$, the Schwarzschild term $-2M/r$ still dominates as $r\rightarrow0$. Therefore, the exponential halo profile softens only the matter-sector contribution, but it does not remove the central Schwarzschild singular behavior. Indeed, using the above expansion, the scalar potential behaves schematically as
\begin{align}
    \mathcal{V}_{\text{scalar}}(r)\simeq \left[\exp\!\left(-4\pi r_s^{2}\rho_s\right)-\alpha-\frac{2M}{r}\right]\left[\frac{\ell(\ell+1)}{r^2}+\frac{2M}{r^3}+\frac{4\pi r_s\rho_s}{r}\exp\!\left(-4\pi r_s^{2}\rho_s\right)+\cdots\right].
\end{align}
Thus, for a nonzero BH mass, the dominant contribution near $r=0$ is singular. This indicates that the model should not be interpreted as a fully regular BH geometry unless the central mass sector is also regularized. From the perturbative viewpoint, the physically relevant region for scattering and quasinormal modes remains the exterior domain outside the event horizon, where the scalar potential forms the effective barrier governing wave propagation.

\section{Conclusion}\label{S6}

In this work, we investigated a Schwarzschild BH surrounded by a Hernquist dark matter halo and supplemented by a constant deformation parameter $\alpha$. The geometry is controlled by the metric function $f(r)$, in which the exponential sector governed by $(r_s,\rho_s)$ describes the extended matter distribution, while $\alpha$ produces a global shift in the spacetime structure. The Schwarzschild solution is recovered when the halo contribution and the deformation parameter are removed.

The analysis of null geodesics showed that the Hernquist halo and the parameter $\alpha$ produce direct corrections to the optical structure of the black hole. In comparison with the Schwarzschild case, the effective potential for photons is modified both in height and radial profile, changing the position of the unstable circular orbit and, consequently, the photon sphere and shadow scale. Physically, $\rho_s$ controls the strength of the matter contribution, whereas $r_s$ determines the radial extension over which the halo affects photon propagation. Thus, denser or more extended halos modify the gravitational trapping region and alter the bending of light. The parameter $\alpha$, on the other hand, changes the asymptotic normalization of the metric function and therefore acts as a global deformation of the optical geometry.

The photon trajectory equation confirms this behavior. Besides the usual Schwarzschild cubic contribution in $u=1/r$, the exponential Hernquist term introduces a non-polynomial radial correction. This correction becomes more relevant in the strong-field region, where photon paths are more sensitive to the structure of the effective potential. Therefore, the parameters $(r_s,\rho_s,\alpha)$ provide a continuous mechanism for deforming the Schwarzschild light trajectories, affecting the lensing pattern, the capture region, and the qualitative structure of null orbits.

The thermodynamic sector also exhibits nontrivial corrections. The horizon condition and the ADM mass acquire explicit dependence on the Hernquist parameters and on $\alpha$, showing that the BH mass is no longer determined only by the horizon radius. The Hawking temperature is modified through the surface gravity, with the exponential matter profile changing the radial gradient of the metric function at the horizon. Generally speaking, the halo contribution tends to reduce the effective thermal emission by smoothing the near-horizon gravitational variation, while $\alpha$ shifts the temperature profile through its global contribution to $A(r)$. The entropy remains given by the Bekenstein-Hawking area law, $S=\pi r_+^2$, indicating that the deformation modifies the thermodynamic response but not the fundamental area scaling.

The Gibbs free energy and heat capacity show that the matter distribution can alter the stability structure of the BH. Particularly, the exponential sector may shift the regions where the heat capacity changes sign or diverges, modifying the transition between locally stable and unstable configurations. Hence, the Hernquist halo does not merely introduce a small correction to the Schwarzschild thermodynamics; it redistributes the stability domains in parameter space, while $\alpha$ acts as an additional control parameter for the global thermodynamic behavior.

Finally, the scalar perturbation analysis demonstrated that the same geometric ingredients also affect the dynamical response of the spacetime. The scalar effective potential depends on both $f(r)$ and $f'(r)$, so the Hernquist halo changes the curvature barrier felt by scalar waves, while $\alpha$ modifies its global normalization. At large distances, the halo contribution appears through an effective mass correction, whereas near the central region the exponential factor approaches a constant. However, for $M\neq0$, the Schwarzschild term remains dominant close to the origin, so the exterior scattering region and the near-horizon potential barrier are the physically relevant domains for scalar-wave propagation.

Our results show that the combined action of $(r_s,\rho_s)$ and $\alpha$ produces consistent deviations from the Schwarzschild geometry in the optical, thermodynamic, and perturbative sectors. The Hernquist halo mainly introduces radial-dependent corrections associated with the surrounding matter distribution, whereas $\alpha$ controls a global deformation of the spacetime. These effects modify photon trapping, light deflection, thermal stability, and scalar-field propagation, providing a useful framework for studying BH observables in environments influenced by extended dark matter distributions.

\section*{Acknowledgments}

F. C. E. Lima would like to express their sincere gratitude to the Conselho Nacional de Desenvolvimento Científico e Tecnológico (CNPq - grant No. 171048/2023-7) and Fundação de Amparo à Pesquisa do Estado de São Paulo (FAPESP - grant No. 2025/05176-7) for their valuable support.

\end{document}